\documentclass[lettersize,journal]{IEEEtran}
\usepackage{amsmath,amsfonts}
\usepackage{algorithmic}
\usepackage{algorithm}
\usepackage{array}
\usepackage[caption=false,font=normalsize,labelfont=sf,textfont=sf]{subfig}
\usepackage{textcomp}
\usepackage{stfloats}
\usepackage{url}
\usepackage{verbatim}
\usepackage{graphicx}
\usepackage{cite}
\usepackage{xcolor}
\usepackage{wasysym}
\usepackage{gensymb}
\usepackage{pifont}

\usepackage{multirow}
\usepackage{titlesec}

\titlespacing\section{0pt}{12pt plus 4pt minus 2pt}{0pt plus 2pt minus 2pt}
\titlespacing\subsection{0pt}{12pt plus 4pt minus 2pt}{0pt plus 2pt minus 2pt}
\titlespacing\subsubsection{0pt}{12pt plus 4pt minus 2pt}{0pt plus 2pt minus 2pt}

\newcommand{\mypara}[1]{\vspace{2pt}\noindent{\textbf{#1}}}

\definecolor{custom_purple}{rgb}{153, 0, 255} 
\newcommand\bs[1]{\textcolor{black}{#1}}

\begin{document}

\title{CVA6 RISC-V Virtualization: Architecture, Microarchitecture, and Design Space Exploration}

\author{\IEEEauthorblockN{
Bruno Sá \IEEEauthorrefmark{1},
Luca Valente \IEEEauthorrefmark{2},
José Martins \IEEEauthorrefmark{1},
Davide Rossi \IEEEauthorrefmark{2}, 
Luca Benini \IEEEauthorrefmark{2} \IEEEauthorrefmark{3},
Sandro Pinto \IEEEauthorrefmark{1}} \\
\IEEEauthorblockA{ 
\IEEEauthorrefmark{1} Centro ALGORTIMI/LASI, Universidade do Minho, Portugal \\ \hspace{2pt}  
\IEEEauthorrefmark{2} DEI, University of Bologna, Italy \hspace{2pt}  
\IEEEauthorrefmark{3} IIS lab, ETH Zurich, Switzerland \\
}
}
\maketitle

\begin{abstract}
Virtualization is a key technology used in a wide range of applications, from cloud computing to embedded systems. Over the last few years, mainstream computer architectures were extended with hardware virtualization support, giving rise to a set of virtualization technologies (e.g., Intel VT, Arm VE) that are now proliferating in modern processors and SoCs.
In this article, we describe our work on hardware virtualization support in the RISC-V CVA6 core. Our contribution is multifold and encompasses architecture, microarchitecture, and design space exploration. In particular, we highlight the design of a set of microarchitectural enhancements (i.e., G-Stage Translation Lookaside Buffer (GTLB), L2 TLB) to alleviate the virtualization performance overhead. We also perform a Design Space Exploration (DSE) and accompanying post-layout simulations (based on 22nm FDX technology) to assess Performance, Power ,and Area (PPA). Further, we map design variants on an FPGA platform (Genesys 2) to assess the functional performance-area trade-off. Based on the DSE, we select an optimal design point for the CVA6 with hardware virtualization support. For this optimal hardware configuration, we collected functional performance results by running the MiBench benchmark on Linux atop Bao hypervisor for a single-core configuration. We observed a performance speedup of up to 16\% (approx. 12.5\% on average) compared with virtualization-aware non-optimized design at the minimal cost of 0.78\% in area and 0.33\% in power. \bs{Finally, all work described in this article is publicly available and open-sourced for the community to further evaluate additional design configurations and software stacks.}
\end{abstract}

\begin{IEEEkeywords}
Virtualization, CVA6, Microarchitecture, TLB, MMU, Design Space Exploration, Hypervisor, RISC-V. 
\end{IEEEkeywords}

\section{Introduction}

Virtualization is a technological enabler used on a large spectrum of applications, ranging from cloud computing and servers to mobiles and embedded systems \cite{Martins2020}. As a fundamental cornerstone of cloud computing, virtualization provides numerous advantages for workload management, data protection, and cost-/power-effectiveness \cite{Heiser2011}. On the other side of the spectrum, the embedded and safety-critical systems industry has resorted to virtualization as a fundamental approach to address the market pressure to minimize size, weight, power, and cost (SWaP-C), while guaranteeing temporal and spatial isolation for certification (e.g., ISO26262) \cite{Bechtel2019,Pinto2019,Cerdeira2022}. Due to the proliferation of virtualization across multiple industries and use cases, prominent players in the silicon industry started to introduce hardware virtualization support in mainstream computing architectures (e.g., Intel Virtualization Technology, Arm Virtualization Extensions, respectively) \cite{Uhlig2005, Arm2018sel2}. 

Recent advances in computing architectures have brought to light a novel instruction set architecture (ISA) named RISC-V \cite{Asanovic2014}. RISC-V has recently reached the mark of 10+ billion shipped cores \cite{nick2022}.
It distinguishes itself from the classical mainstream by providing a free and open standard ISA, featuring a modular and highly customizable extension scheme that allows it to scale from small microcontrollers up to supercomputers \cite{gautschi_2017,Valente2022,Rossi2022,Garofalo2022,zaruba2019,Chen2020,Ficarelli2022}. The RISC-V privileged architecture provides hardware support for virtualization by defining the Hypervisor extension \cite{RISCV2022}, ratified in Q4 2021.

Despite the Hypervisor extension ratification, as of this writing, there is no RISC-V silicon with this extension on the market\footnote{SiFive, Ventana, and StarFive have announced RISC-V CPU designs with Hypervisor extension support, but we are not aware of any silicon available on the market yet.}. There are open-source hypervisors with upstream support for the Hypervisor extension, i.e., Bao \cite{Martins2020}, Xvisor \cite{Patel2015}, KVM \cite{Zhao2020}, and seL4 \cite{Heiser2020} (and work in progress in Xen \cite{Hwang2008} and Jailhouse \cite{Ramsauer2022}). However, to the best of our knowledge, there are just a few hardware implementations deployed on FPGA, which include the Rocket chip \cite{sa2021} and NOEL-V \cite{Andersson2020} (and soon SHAKTI and Chromite \cite{Gala2016}). Notwithstanding, no existing work has (i) focused on understanding and enhancing the microarchitecture for virtualization and (ii) performed a design space exploration (DSE) and accompanying power, performance, area (PPA) analysis.

This work describes the architectural and microarchitectural support for virtualization in an open-source RISC-V CVA6-based \cite{zaruba2019} (64-bit) SoC. At the architectural level, the implementation is compliant with the Hypervisor extension (v1.0) \cite{RISCV2022} and includes the implementation of the RISC-V timer (Sstc) extension \cite{scheid2021} as well. At the microarchitectural level, we modified the vanilla CVA6 microarchitecture to support the Hypervisor extension and proposed a set of additional extensions/enhancements to reduce the hardware virtualization overhead: (i) a dedicated second stage \bs{Translation Lookaside Buffer (TLB)} coupled to the Page Table Walker (PTW) (i.e., G-Stage TLB (GTLB) in our lingo), and (ii) a second level TLB (L2 TLB). We also present and discuss a comprehensive design space exploration on the microarchitecture. We first evaluate 23 (out of 288) hardware designs deployed on FPGA (Genesys 2) and assess the impact on functional performance (execution cycles) and hardware. Then, we elect 7 designs and analyze them in depth with post-layout simulations of implementations in 22nm FDX technology. 

\bs{We ran the MiBench (automotive subset) benchmarks for the DSE evaluation to assess functional performance. The virtualization-aware, non-optimized CVA6 implementation served as our baseline configuration. The software setup encompassed a single Linux Virtual Machine (VM) running atop Bao hypervisor for a single-core design. We measured the performance speedup of the hosted Linux VM relative to the baseline configuration. Results from the DSE exploration demonstrated that the proposed microarchitectural extensions could achieve a functional performance speedup up to 19\% (e.g., for the \textit{susanc} small benchmark); however, in some cases at the cost of a non-negligible increase in area and power. Thus, results from the PPA analysis show that: (i) the Sstc extension has negligible impact on power and area; (iii) the GTLB increases the overall area in less than 1\%; and (iv) the L2 TLB introduces a non-negligible 8\% increase in area in some configurations. As a representative, well-balanced configuration, we selected the CVA6 design with Sstc support and a GTLB with 8 entries. For this specific hardware configuration, we observed a performance speedup of up to 16\% (approx. 12.5\% on average) at the cost of 0.78\% in area and 0.33\% in power. To the best of our knowledge, this paper reports the first public work on a complete DSE evaluation and PPA analysis for a virtualization-enhanced RISC-V core.}

\bs{To summarize, with this work, we make the following contributions. Firstly, we provide hardware virtualization support for the CVA6 core, which was completely absent in the vanilla implementation. In particular, we implement the RISC-V Hypervisor extension (v1.0) and design a set of (virtualization-oriented) microarchitectural enhancements to the Nested Memory Management Unit (Section \ref{sec:cva6_hyp}). To the best of our knowledge, no work or study describes and discusses microarchitectural extensions to improve the hardware virtualization support in a RISC-V core. Second, we perform a DSE encompassing dozens of design configurations. This DSE includes trade-offs on parameters from three different microarchitectural components (L1 TLB, GTLB, L2 TLB) and respective impact on functional performance and hardware costs (Section \ref{sec:design-space-exploration}). Finally, we conduct post-layout simulations on a few elected design configurations to assess a PPA analysis (Section \ref{ppa-analysis}). All contributions described in this manuscript are open source \footnote{https://github.com/minho-pulp/cva6/} and available to the RISC-V community to foster collaboration and enable contributions with further extensions, optimizations, and testing/verification activities. The Hypervisor extension is now going towards formal upstream into the CVA6 main repository\footnote{https://github.com/minho-pulp/cva6/tree/feat/hyp-upstream}. Our goal is to democratize virtualization for the next-generation CVA6-based SoCs.}

\section{Background}

\bs{This section covers the background related to RISC-V technology (ISA, extensions, and cores) and virtualization. To improve readability, we also provide a table of abbreviations and terminologies used throughout the article (see Table \ref{tab:abb}).}

\subsection{Virtualization Technology}
\bs{Virtualization is the de facto technology that consolidates and isolates multiple non-related software stacks onto the same hardware platform by partitioning and multiplexing hardware resources (e.g., CPUs, memory) between multiple virtual machines. The virtual machine monitor (VMM) or hypervisor is the software layer that implements virtualization. The software executing in the VM is referred to as a guest, typically an operating system (OS), i.e., guest OS.}

\subsection{RISC-V Privileged Specification}
The RISC-V privileged instruction set architecture (ISA) \cite{RISCV2022} divides its execution model into 3 privilege levels: (i) \textit{machine mode} (M-mode) is the most privileged level, hosting the firmware which implements the supervisor binary interface (SBI) (e.g., OpenSBI); (ii) \textit{supervisor mode} (S-Mode) runs Unix type operating systems (OSes) that require virtual memory management; (iii) \textit{user mode} (U-Mode) executes userland applications. The modularity offered by the RISC-V ISA seamlessly allows for implementations at distinct design points, ranging from small embedded platforms with just M-mode support to fully blown server class systems with M/S/U.

\subsection{RISC-V Virtualization} \label{subsec:risc-v-virt}

\begin{figure}[t]
    \centering
    \includegraphics[width=0.47\textwidth,clip]{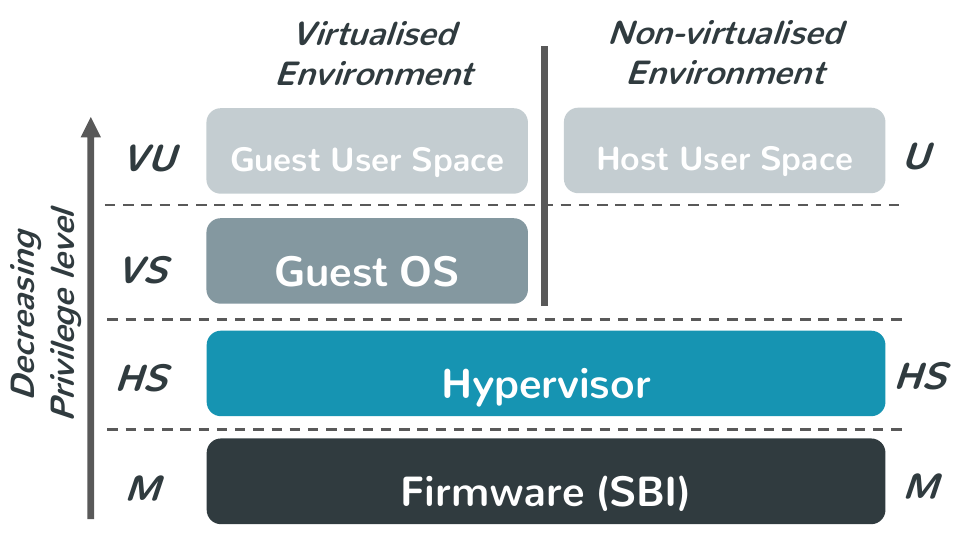}
    \caption{RISC-V privilege levels: machine (M), hypervisor-extended supervisor (HS), virtual supervisor (VS), and virtual user (VU).}
    \label{fig:exec_modes}
    \vspace{-5mm}
\end{figure}

Unlike other mainstream ISAs, the RISC-V privileged architecture was designed from the initial conception to be classically virtualizable \cite{Popek1974}. So although the ISA, per se, allows the straightforward implementation of hypervisors resorting, for example, to classic virtualization techniques (e.g., trap-and-emulation and shadow page tables), it is well understood that such techniques incur a prohibitive performance penalty and cannot cope with current embedded real-time virtualization requirements (e.g., interrupt latency) \cite{sa2021}. Thus, to increase virtualization efficiency, the RISC-V privileged architecture specification introduced hardware support for virtualization through the (optional) \textit{Hypervisor} extension \cite{RISCV2022}. \bs{The following paragraphs provide a high-level overview of the RISC-V Hypervisor extension specification.}

\begin{table*}[!t]
\caption{\bs{Table of abbreviations, RISC-V terminologies and RISC-V Control Status Registers and Instructions.}}
\center
\resizebox{1\textwidth}{!}{
\begin{tabular}{ll}
\hline
\multicolumn{1}{|l|}{\textbf{Name}}            & \multicolumn{1}{l|}{\textbf{Definition}}                                                                                                                                                                                                                                                                                            \\ \hline
                                               &                                                                                                                                                                                                                                                                                                                                                                                \\ \hline
\multicolumn{2}{|l|}{\textbf{Abbreviations}}                                                                                                                                                                                                                                                                                                                                                                                      \\ \hline
\multicolumn{1}{|l|}{PTW}                                                                    & \multicolumn{1}{l|}{Page Table Walker. The PTW is responsible for transversing the page table and converting a virtual address into a physical address.}                                                                                                                                                \\ \hline
\multicolumn{1}{|l|}{DSE}                                                       & \multicolumn{1}{l|}{Design Space Exploration. Functional evaluation carried out for multiple FPGA design configurations.}                                                                                                                                       \\ \hline
\multicolumn{1}{|l|}{PPA}                                                          & \multicolumn{1}{l|}{Power, Performance, Area. Post-layout silicon power, performance, and area analysis.}                                                                                                                                                       \\ \hline
\multicolumn{1}{|l|}{VM}                                                                         & \multicolumn{1}{l|}{Virtual Machine. The execution environment (e.g., OS or application) running atop a hypervisor.}                                                                                                                                                                                   \\ \hline
\multicolumn{1}{|l|}{Nested-MMU}                                              & \multicolumn{1}{l|}{\begin{tabular}[c]{@{}l@{}}Nested Memory Management Unit. Using two translation stages, convert a guest's virtual address into a host's physical address. \\ The first is controlled by the guest OS and the second by the hypervisor.\end{tabular}} \\ \hline
\multicolumn{1}{|l|}{TLB}                                                         & \multicolumn{1}{l|}{Translation Lookaside Buffer. Microarchitectural element used to cache translations from virtual addresses to physical addresses.}                                                                                                                                                                          \\ \hline
\multicolumn{1}{|l|}{VS-Mode}                                                  & \multicolumn{1}{l|}{Virtual Supervisor Mode. Privilege mode in RISC-V, where a guest OS runs when virtualized.}                                                                                                                                                                                                \\ \hline
\multicolumn{1}{|l|}{HS-Mode}                                         & \multicolumn{1}{l|}{Hypervisor-extended Supervisor Mode. Privilege mode in RISC-V, where the hypervisor runs.}                                                                                                                                                                                                             \\ \hline
\multicolumn{1}{|l|}{GTLB}                                       & \multicolumn{1}{l|}{G-stage Translation Lookaside Buffer. TLB that stores G-stage only translation, i.e., guest physical addresses into host physical addresses.}                                                                                                                          \\ \hline
                                               &                                                                                                                                                                                                                                                                                                                                                                                 \\ \hline
\multicolumn{2}{|l|}{\textbf{RISC-V Terminologies}}                                                                                                                                                                                                                                                                                                                                                                               \\ \hline
\multicolumn{1}{|l|}{CSR}                                                      & \multicolumn{1}{l|}{Control Status Registers. RISC-V CPU control registers.} 
\\ \hline
\multicolumn{1}{|l|}{VS-Stage}                                  & \multicolumn{1}{l|}{RISC-V terminology for a first translation stage. Converts guest virtual addresses into guest physical addresses.}                                                                                                                                                                                                 \\ \hline
\multicolumn{1}{|l|}{G-Stage}                                 & \multicolumn{1}{l|}{RISC-V terminology for a second translation stage. Converts guest physical addresses into host physical addresses.}                                                                                                                                                                                                 \\ \hline
\multicolumn{1}{|l|}{ASID}                                                     & \multicolumn{1}{l|}{Address Space Identifier. Processes have a unique identifier used to tag translations on the MMU (e.g., on TLB entries).}                                                                                                                                                               \\ \hline
\multicolumn{1}{|l|}{VMID}                                                       & \multicolumn{1}{l|}{Virtual Machine Space Identifier. Guest virtual machines have a unique identifier used to tag translations on the MMU.}                                                                                                                                                                         \\ \hline
\multicolumn{1}{|l|}{PTE}                                                                 & \multicolumn{1}{l|}{Page Table Entry. Specific entry of a page table used to convert virtual addresses into physical addresses.}                                                                                                                                                       \\ \hline
                                               &                                                                                                                                                                                                                                                                                                                                                                                 \\ \hline
\multicolumn{2}{|l|}{\textbf{RISC-V Control Status Registers and Instructions}}                                                                                                                                                                                                                                                                                                                                                   \\ \hline
\multicolumn{1}{|l|}{\textit{stimecmp}}                               & \multicolumn{1}{l|}{Supervisor Time Comparator Register. Generates timer interrupts at HS/S-mode. Only available if the RISC-V Sstc extension is implemented.}                                                                                                                                                            \\ \hline
\multicolumn{1}{|l|}{\textit{vstap}}     & \multicolumn{1}{l|}{Virtual Supervisor Guest Address Translation and Protection Register. Controls the VS-stage address translation and protection.}                                                                                                                                                                                                       \\ \hline
\multicolumn{1}{|l|}{\textit{vstimecmp}}      & \multicolumn{1}{l|}{Virtual Supervisor Time Comparator Register. Generates timer interrupts at VS-mode. Only available if the RISC-V Sstc extension is implemented.}                                                                                                                                                              \\ \hline
\multicolumn{1}{|l|}{\textit{hgeie and hgeip}} & \multicolumn{1}{l|}{Hypervisor Guest External Enable and Pending Registers. Deliver interrupts directly to the guest if the interrupt controller supports virtualization.}                                                                                                                                                           \\ \hline
\multicolumn{1}{|l|}{\textit{hvencfg}}                      & \multicolumn{1}{l|}{Hypervisor Virtual Environment Control Register. Controls hypervisor access (enabled or disabled) to optional platform-specific extensions (e.g., Sstc extension).}                                                                                                                                               \\ \hline
\multicolumn{1}{|l|}{\textit{hgatp}}           & \multicolumn{1}{l|}{Hypervisor Guest Address Translation and Protection Register. G-stage root table pointer and respective translation-specific configuration fields.}                                                                                                                                                                  \\ \hline
\multicolumn{1}{|l|}{\textit{menvcfg}}                               & \multicolumn{1}{l|}{Machine Virtual Environment Control Register. Controls firmware access (enabled or disabled) to optional platform-specific extensions (e.g., Sstc extension).}                                                                                                                                                  \\ \hline
\multicolumn{1}{|l|}{\textit{hfence}}        & \multicolumn{1}{l|}{\begin{tabular}[c]{@{}l@{}}Hypervisor Fence Instruction. Flush cached translations on TLBs and other related structures. \\ The hfence.vvma flushes VS-Stage and G-Stage translations. The hfence.gvma flushes only G-Stage translations.\end{tabular}}                              \\ \hline
\end{tabular}
}
\label{tab:abb}
\vspace{-5mm}
\end{table*}

\mypara{Privilege Levels.} As depicted in Figure \ref{fig:exec_modes}, the RISC-V Hypervisor extension execution model follows an orthogonal design where the \textit{supervisor mode} (S-mode) is modified to an \textit{hypervisor-extended supervisor mode} (HS-mode) well-suited to host both type-1 or type-2 hypervisors\footnote{The main difference between type-1 (or baremetal) and type-2 (or hosted) hypervisors is that a type-1 hypervisor runs directly on the hardware (e.g., Xen) while a type-2 hypervisor runs atop an operating system (e.g., VMware).}. Additionally, two new privileged modes are added and can be leveraged to run the guest OS at \textit{virtual supervisor mode} (VS-mode) and \textit{virtual user mode} (VU-mode). 

\mypara{Two-stage Address Translation.} \bs{The Hypervisor extension also defines a second translation stage (\textit{G-stage} in RISC-V lingo) to virtualize the guest memory by translating guest-physical addresses into host-physical addresses. The HS-mode operates like S-mode but with additional hypervisor registers and instructions to control the VM execution and \textit{G-stage} translation. For instance, the \textit{hgatp} register holds the \textit{G-stage} root table pointer and translation-specific configuration fields.}

\mypara{Hypervisor Control and Status Registers (CSRs).} Each VM running in VS-mode has its own control and status registers (CSRs) that are shadow copies of the S-mode CSRs. These registers can be used to determine the guest execution state and perform VM switches. To control the virtualization state, a specific flag called \textit{virtualization mode} (V bit) is used. When V=1, the guest is executing in VS-mode or VU-mode, normal S-mode CSRs accesses are actually accessing the VS-mode CSRs, and the \textit{G-stage} translation is active. Otherwise, if V=0, normal S-mode CSRs are active, and the \textit{G-stage} is disabled. To ease guest-related exception trap handling, there are guest-specific traps, e.g., guest page faults, VS-level illegal exceptions, and VS-level \textit{ecalls} (a.k.a. hypercalls).

\mypara{Hypervisor Instructions.} \bs{The Hypervisor extension defines a set of hypervisor-related instructions to increase the virtualization efficiency. These instructions encompass: (i) hypervisor load/store instructions used to access guest memory with the exact translation and permissions rights (RWX) as the guest; and (ii) hypervisor fence instructions synchronization mechanisms to flush guest-related cached translation structures (e.g., TLBs) during context switching operations.}

\subsection{Nested Memory Management Unit (Nested-MMU)}

The MMU is a hardware component responsible for translating virtual memory references to physical ones while enforcing memory access permissions. The OS controls the MMU by assigning virtual address space to each process and managing the MMU translation structures to translate virtual addresses into physical addresses correctly. \bs{On a virtualized system, the MMU provides another layer of translation and protection controlled by the hypervisor. In this case, the MMU can translate from guest-virtual addresses to guest-physical addresses and from guest-physical addresses into host-physical addresses. This new feature is referred to as nested-MMU. The RISC-V ISA supports the nested-MMU through a new stage of translation that converts guest-physical addresses into host-physical addresses, denoted G-stage. The guest VM takes control over the first stage of translation (VS-stage in RISC-V lingo), while the hypervisor assumes control over the second one (G-stage). Originally, the RISC-V privileged specification defines that a virtual address is converted into a physical address by traversing a multi-level radix-tree table using one of four different topologies: (i) \textit{Sv32} for 32 virtual address spaces with a 2-level hierarchy tree; (ii) \textit{Sv39} for 39-bit virtual address spaces with a 3-level tree; (iii) \textit{Sv48} for 48-bit VAS and 4-level tree; and (iv) \textit{Sv57} for 57-bit virtual address spaces and 5-level tree. Each level holds a pointer to the following table (non-leaf entry) or the final translation (leaf entry). This pointer and permissions are stored in a 64-bit (RV64) or 32-bit (RV32) width page table entry (PTE). Note that RISC-V splits the virtual address into 4KiB page sizes, but since each level can either be a leaf or non-leaf, it supports superpages to reduce the TLB pressure, e.g., Sv39 supports 4KiB, 2MiB, and 1GiB page sizes.}

\subsection{RISC-V "stimecmp/vstimecmp" Extension (Sstc)}

\bs{The RISC-V Sstc extension \cite{scheid2021} aims to enhance supervisor mode with its timer interrupt facility, thus eliminating the large overheads for emulating S/HS-mode timers and timer interrupt generation up in M-mode. The Sstc extension also adds a similar facility to the Hypervisor extension for VS-mode. To enable direct control over timer interrupts in the HS/S-mode and VS-mode, the Sstc encompasses two additional comparators: (i) \textit{stimecmp} register to generate S/HS-mode timer interrupts and (ii) \textit{vstimecmp} register to generate VS-mode timer interrupts. Whenever the value of the system clock \textit{time} counter register is greater than the value on the comparators, a timer interrupt is generated.} For a complete overview of the RISC-V timer architecture and a discussion on why the classic RISC-V timer specification incurs a significant performance penalty, we refer the interested reader to \cite{sa2021}.

\begin{figure*}[!t]
    \centering
    \includegraphics[width=0.92\textwidth,clip]{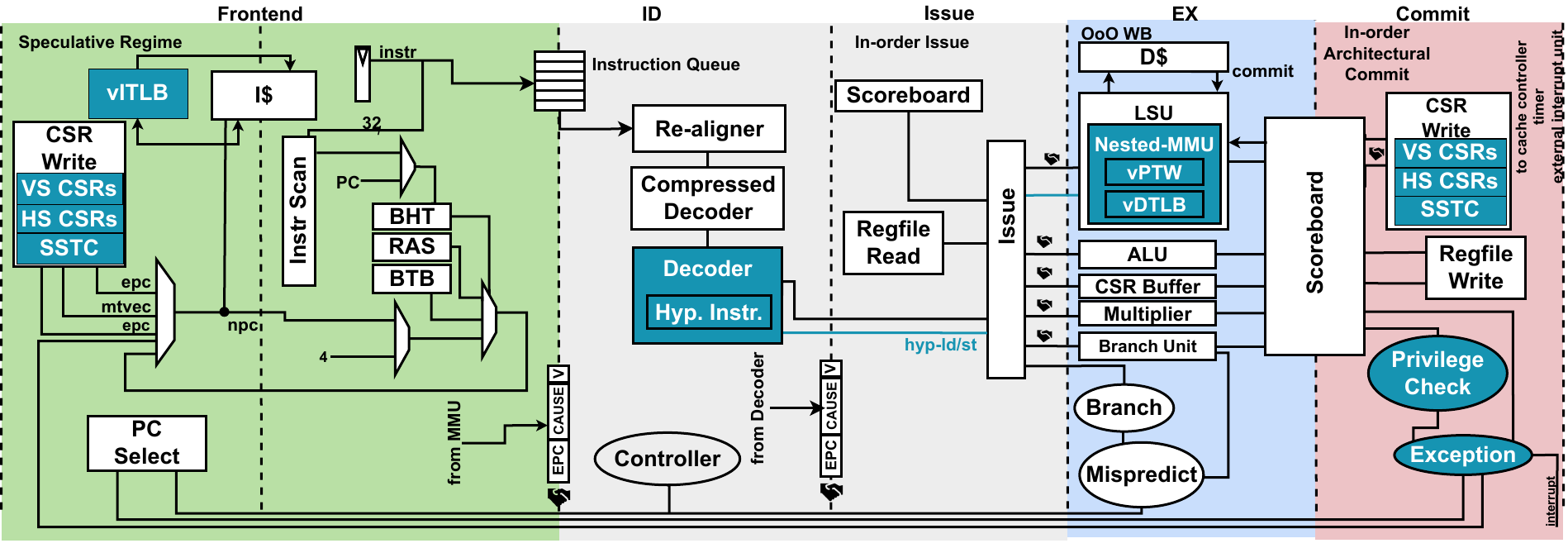}
    \caption{\bs{CVA6 core microarchitecture featuring the RISC-V hypervisor extension. Major microarchitectural changes to the core functional blocks (e.g., Decoder, PTW, TLBs, and CSRs) are highlighted in blue. Adapted from \cite{zaruba2019}.}}
    \label{fig:cva6-h-over}
    \vspace{-5mm}
\end{figure*}

\subsection{CVA6}

CVA6 is an application class RISC-V core that implements both the RV64 and RV32 versions of RISC-V ISA \cite{zaruba2019}. The core fully supports the 3 privilege execution modes M/S/U-modes and provides hardware support for memory virtualization by implementing a MMU, making it suitable for running a fully-fledged OS such as Linux. Recent additions also include a energy-efficient Vector Unit co-processor \cite{cavalcante_2020}. Internally, the CVA6 microarchitecture encompasses a 6-stage pipeline, single issue, with an out-of-order execution stage and 8 PMP entries. The MMU has separate TLBs for data and instructions and the PTW implements the \textit{Sv39} and \textit{Sv32} translation modes as defined by the privileged specification \cite{RISCV2022}.

\section{CVA6 "From Scratch" Virtualization Support} \label{sec:cva6_hyp}

\bs{This section reports our work in adding hardware support for virtualization in the CVA6 core (compliant with the RISC-V Hypervisor extension v1.0). To the best of our knowledge, this work is the first initiative to add hardware virtualization support in the CVA6 core. The following section divides into seven logical subsections: (i) Subsection \ref{subsec:hyp_virt_exec}, \ref{subsec:hyp-ld-st}, \ref{subsec:nested-ptw} and \ref{subsec:nested-tlb} describes the modifications/additions to CVA6 microarchitecture enabling the Hypervisor extension support, specifically to the core registers and the MMU subsystem; (ii) Subsection \ref{subsec:gtlb} and \ref{subsec:l2-tlb} presents the microarchitectural enhancements designed in the MMU subsystem to increase virtualization efficiency; and (iii) Subsection \ref{subsec:sstc} describes how we extend the CVA6 timer with the standard RISC-V Sstc extension to improve the timer management infrastructure. Figure \ref{fig:cva6-h-over} illustrates a high-level overview of the modifications/additions performed in the CVA6 microarchitecture.}

\subsection{Hypervisor and Virtual Supervisor Execution Modes} \label{subsec:hyp_virt_exec}

As previously described (refer to Section \ref{subsec:risc-v-virt}), the Hypervisor extension specification extends the S-mode into the HS-mode and adds two extra orthogonal execution modes, denoted VS-mode and VU-mode. To add support for these new execution modes, we have extended/modified some of the CVA6 core functional blocks, in particular, the \textit{CSR} and \textit{Decode} modules. As illustrated by Figure \ref{fig:cva6-h-over}, the hardware virtualization architecture logic encompasses five building blocks: (i) VS-mode and HS-mode CSRs access logic and permission checks; (ii) exceptions and interrupts triggering and delegation; (iii) trap entry and exit; (iv) hypervisor instructions decoding and execution; and (v) nested-MMU translation logic. The \textit{CSR module} was extended to implement the first three building blocks that comprise the hardware virtualization logic, specifically: (i)  HS-mode and VS-mode CSRs access logic (read/write operations); (ii) HS and VS execution mode trap entry and return logic; and (iii) a fraction of the exception/interrupt triggering and delegation logic from M-mode to HS-mode and/or to VS/VU-mode (e.g., reading/writing to \textit{vsatp} CSR triggers an exception in VS-mode \bs{when not allowed by the hypervisor}). The \textit{Decode} module was modified to implement hypervisor instructions decoding (e.g., hypervisor load/store instructions and memory-management fence instructions) and all VS-mode related instructions execution access exception triggering. 

We refer readers to Table \ref{tab:h-feat}, which presents a summary of the features that were fully and partially implemented. We have implemented all mandatory features of the ratified specification (v1.0); however, we still left some optional features as partially implemented due to the dependency on upcoming or newer extensions. For example, \textit{hvencfg} bits depend on \textit{Zicbom} \cite{kruckemyer2022} (cache block management operations); \textit{hgeie} and \textit{hgeip} depends on the Advanced Interrupt Architecture (AIA) \cite{hauser2022}; and \textit{hgatp} depends on virtual address spaces not currently supported in the vanilla CVA6.

\begin{table}[t]
    \caption{Hypervisor Extension features implemented in the CVA6 core: \CIRCLE \hspace{1pt} fully-implemented; \LEFTcircle \hspace{1pt} partially implemented.}
    \center
    \begin{tabular}{|c|l|c|}
    \hline
    \multirow{11}{*}{CSRs} 
     & hstatus/mstatus   &  \CIRCLE  \\ \cline{2-3}   
     & hideleg/hedeleg/mideleg &  \CIRCLE  \\ \cline{2-3}
     & hvip/hip/hie/mip/mie    &   \CIRCLE \\ \cline{2-3}  
     & hgeip/hgeie   &   \LEFTcircle  \\ \cline{2-3}         
     & hcounteren   &   \CIRCLE  \\ \cline{2-3}  
     & htimedelta   &  \CIRCLE \\ \cline{2-3}  
     & henvcfg      &  \LEFTcircle \\ \cline{2-3} 
     & mtval2/htval   &   \CIRCLE  \\ \cline{2-3}        
     & mtinst/htinst   & \CIRCLE  \\ \cline{2-3}  
     & hgapt   &  \LEFTcircle  \\ \cline{2-3}          
     &  \begin{tabular}[c]{@{}l@{}}vsstatus/vsip/vsie/vstvec/vsscratch\\vsepc/vscause/vstval/vsatp \end{tabular} &  \CIRCLE \\ \hline \hline   
    \multirow{2}{*}{Intructions}
        & hlv/hlvx/hsv & \CIRCLE \\ \cline{2-3} 
        & hfence.vvma/gvma & \CIRCLE \\ \hline \hline
    \multirow{6}{*}{Exceptions \& Interrupts}
        & Environment call from VS-mode & \CIRCLE  \\ \cline{2-3} 
        & Instruction/Load/Store guest-page fault & \CIRCLE  \\ \cline{2-3} 
        & Virtual instruction &  \CIRCLE \\ \cline{2-3} 
        & \begin{tabular}[c]{@{}l@{}} Virtual Supervisor sw/timer/external\\interrupts\end{tabular} & \CIRCLE  \\ \cline{2-3} 
        & Supervisor guest external interrupt &  \CIRCLE \\ \hline
    \end{tabular}
    \label{tab:h-feat}
    \vspace{-5mm}
    \end{table}

\subsection{Hypervisor Load/Store Instructions} \label{subsec:hyp-ld-st}

The hypervisor load/store instructions (i.e., \textit{HLV}, \textit{HSV}, and \textit{HLVX}) provide a mechanism for the hypervisor to access the guest memory while subject to the exact translation and permission checks as in VS-mode or VU-mode. These instructions change the translation settings at the instruction granularity level, forcing a complete swap of privilege level and translation applied at every hypervisor load/store instruction execution. The implementation encompasses the addition of a signal (identified as \textit{hyp\_ld/st} in Figure \ref{fig:cva6-h-over}) to the CVA6 pipeline that travels from the decoding to the load/store unit in the execute stage. This signal is then fed into the MMU that performs (i) all necessary context switches (i.e., enables the \textit{hgatp} and \textit{vstap} CSRs), (ii) enables the virtualization mode, and (iii) changes execution mode.

\subsection{Nested Page-Table Walker} \label{subsec:nested-ptw}

One of the major functional blocks of an MMU is the PTW. Fundamentally, the PTW is responsible for partitioning a virtual address accordingly to the specific topology and scheme (e.g., \textit{Sv39}) and then translating it into a physical address using the memory page tables structures. The Hypervisor extension specifies a new translation stage (G-stage) used to translate guest-physical addresses into host-physical addresses. Our implementation supports \textit{Bare} translation mode (no G-stage) and \textit{Sv39x4}, which defines a 41-bit width maximum guest physical address space. \bs{As of this writing, the CVA6 only supported the \textit{Sv39} scheme for the 64-bit core, so we extended the implementation to support the \textit{Sv39x4}}.

We extended the existing finite state machine (FSM) used to translate virtual addresses to physical addresses and added only a new control state to keep track of the current translation stage and assist the context switching between VS-Stage and G-Stage translations. With the G-stage in situ, it is mandatory to translate: (i) the resulting guest-physical address from the VS-Stage translation and (ii) all memory accesses made during the VS-Stage translation walk. To accomplish that, we identify three stages of translation that can occur during a PTW iteration: (i) \textit{VS-Stage} - the PTW current state is translating a guest-virtual address into a guest-physical address; (ii) \textit{G-Stage Intermed} - the PTW current state is translating memory access made from the VS-Stage during the walk to host-physical address; and (iii) \textit{G-Stage Final} - the PTW current state is translating the final output address from VS-Stage into a host-physical address. It is worth noting that if MMU is configured in Bare mode for the G-stage translation, we perform a standard S/VS-Stage translation. Once the nested walk completes, the PTW updates the TLB with the final PTE from VS-stage and G-stage, alongside the current Address Space Identifier (ASID) and the Virtual-Machine Identifier (VMID). \bs{ VMID tags each translation to a specific VM.} One implemented optimization consists of storing the translation page size (i.e., 4KiB, 2MiB, and 1GiB) for both VS- and G-stages into the same TLB entry and permissions access bits for each stage.

\subsection{Virtualization-aware TLBs (vTLB)} \label{subsec:nested-tlb}

The CVA6 MMU microarchitecture has two small, fully associative TLB: data (L1 DTLB) and instructions (L1 ITLB). Both TLBs support a maximum of 16 entries and fully implement the flush instructions, i.e., \textit{sfence}, including filtering by ASID and virtual address. To support nested translation, we modified the L1 DTLB and ITLB microarchitecture to support two translation stages, including access permissions and VMIDs. Each TLB entry holds both VS-Stage and G-Stage PTE and respective permissions. The lookup logic is performed using the merged final translation size from both stages, i.e., if the VS-stage is a 4KiB and the G-stage is a 2MiB translation, the final translation would be a 4KiB. This is probably one of the major drawbacks of having both the VS-stage and G-stage stored together. For instance, hypervisors supporting superpages/hugepages use the 2MiB page size to optimize the MMU performance. Although this significantly reduces the PTW walk time, if the guest uses a 4KiB page size, the TLB lookup would not benefit from superpages since the translation would be stored as a 4KiB page size translation. \bs{A possible alternative would be having separated TLBs for each stage. We argue that this solution would have three major drawbacks: (i) negative impact on performance as it would be possible to search in G-Stage TLB after the VS-Stage TLB lookup (i.e., increase the TLB hit time penalty by a factor of 2); (i) less hardware reuse, i.e., the G-Stage TLB is only used when a guest is running and G-stage is active; and (iii) more impact on the area and energy, the CVA6 TLBs are fully combinational circuits with a significant impact on the area, energy, and timing. For all the above reasons, we decided to keep the VS-stage and G-stage translations in a single TLB entry.}. Finally, the TLB also supports VMID tags allowing hypervisors to perform a more efficient TLB management using per-VMID flushes and avoiding full TLB flush on a VM context switch. Finally, the TLB also allows flushes by guest physical address, i.e., hypervisor fence instructions (\textit{hfence.vvma/gvma}) are fully supported.

\begin{figure}[!t]
    \centering
    \includegraphics[width=0.48\textwidth,clip]{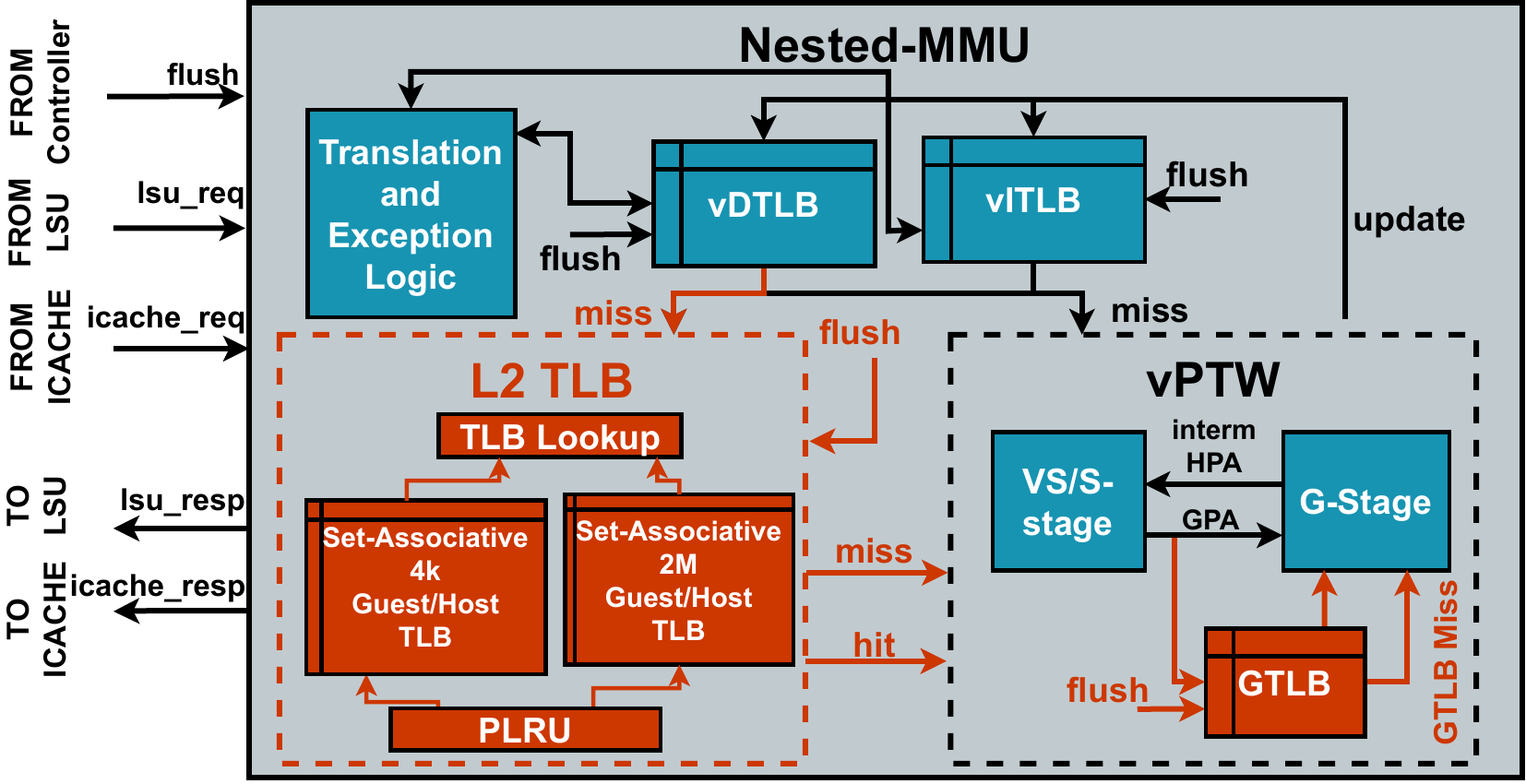}
    \caption{\bs{MMU microarchitectural enhancements: high-level overview. Additions to the
original MMU design highlighted in orange.}}
    \label{fig:mmu-opt}
    \vspace{-5mm}
\end{figure}

\subsection{Microarchitectural extension \#1 - GTLB} \label{subsec:gtlb}
\begin{figure}[t]
    \centering
    \includegraphics[width=0.4\textwidth,clip]{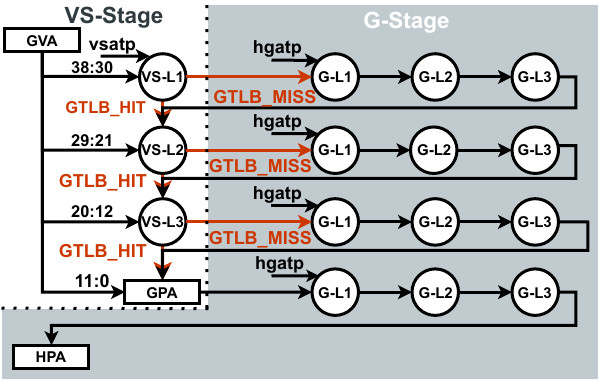}
    \caption{\bs{RISC-V nested-MMU walk example for the SV39x4 scheme.}}
    \label{fig:nested-mmu-gtlb}
    \vspace{-5mm}
\end{figure}

A full nested table walks for the \textit{Sv39} scheme can take up to 15 memory accesses, five times more than a standard S-stage translation. This additional burden imposes a higher TLB miss penalty, resulting in a (measured) overhead of up to 13\% on the functional performance (execution cycles) (comparing the baseline virtualization implementation with the vanilla CVA6). To mitigate this, we have extended the CVA6 microarchitecture with a \bs{GTLB} in the nested-PTW module to store intermediate GPA to HPA translations, i.e., VS-Stage non-leaf PTE guest physical address to host physical addresses translation. Figure \ref{fig:mmu-opt} illustrates the modifications required to integrate the GTLB in the MMU microarchitecture. The GTLB structure aims at accelerating VS-stage translation by skipping each nested translation during the VS-stage page table walk. Figure \ref{fig:nested-mmu-gtlb} presents a 4KiB page size translation process featuring a GTLB in the PTW. Without the GTLB, each time the guest forms the non-leaf physical address PTE pointer during the walk, it needs: (i) to translate it via G-stage, (ii) read the next level PTE value from memory, and (iii) resume the VS-stage translation. When using superpages (2MiB or 1GiB), there are fewer translation walks, reducing the performance penalty. With a simple hardware structure, it is possible to mitigate such overheads by keeping those G-stage translations cached while avoiding unnecessary cache pollution and nondeterministic memory accesses.

\mypara{GTLB microarchitecture.}  The GTLB follows a fully-associative design with support for all \textit{Sv39} superpage sizes (4KiB, 2MiB, and 1GiB). Each entry is searched in parallel during the TLB lookup, and each translation can be stored on any TLB entry. The replacement policy evicts the least recently used entry using a pseudo-least recently used (PLRU) algorithm already implemented on the CVA6 L1 TLBs. The maximum number of entries that GTLB can hold simultaneously ranges from 8 to 16. The flush circuit fully supports hypervisor fence instruction (\textit{HFENCE.GVMA}), including filtering flushed TLB entries by VMIDs and virtual address. It is worth mentioning that the GTLB implementation reused the already-in-place L1 TLB design and modified it to store only G-stage-related translations. We map TLB entries to flip-flops due to their reduced number.

\subsection{Microarchitectural extension \#2 - L2 TLB} \label{subsec:l2-tlb}
Currently, the vanilla CVA6 MMU features only a small separate L1 instruction and data TLBs, shared by guests and the hypervisor. The TLB can merge VS and G stage translations in one single entry, where the stored translation page size must be the minimum of the two stages. This may improve hardware reuse and save some additional search cycles, as no separated TLBs for each stage are required. However, as explained in Subsection \ref{subsec:nested-tlb}, one major drawback of this approach is that if the hypervisor or the guest uses superpages, the TLB would not benefit from them if the pages differ in size, i.e., there would be less TLB coverage than expected. This would result in more TLB misses and page-table walks and, naturally, a significant impact on the overall performance. To deal with the increased TLB coverage and TLB miss penalty caused by G-stage overhead, as well as with the inefficiency arising from the mismatch between translation sizes, we have augmented the CVA6 MMU with a large set-associative private unified L2 TLB as illustrated in Figure \ref{fig:mmu-opt}.

\mypara{L2 TLB microarchitecture.} The TLB stores each translation size in different structures to simplify the look-up logic. The L2 TLB follows a set-associative design with translation tags and data stored in SRAMs. As implemented in the GTLB, the replacement policy is also PLRU. To speed up the search logic, the L2 TLB look-ups and the PTW execute in parallel, thus not affecting the worst-case L1 TLB miss penalty and optimizing the L2 miss penalty. Moreover, the L2 TLB performs searches in parallel for each page size (4KiB or 2MIB), i.e., each page size translation is stored on different hardware structures with independent control and storage units. SFENCE and HFENCE instructions are supported, but a flush signal will flush all TLB entries, i.e., no filtering by VMIDs or ASIDs. We implemented the L2 TLB controller \bs{using a 4 state FSM}. First, in \textit{Flush} state, the logic encompasses walking through the entire TLB and invalidating all TLB entries. Next, in the \textit{Idle} state, the FSM waits for a valid request or update signal from the PTW module. Third, in the \textit{Read} state, the logic performs a read and tag comparison on the TLB set entries. If there is a hit during the look-up, it updates the correct translation and hit signals to the PTW. If there is no look-up hit, the FSM goes to the IDLE state and waits for a new request. Finally, in the \textit{Update} state, we update the TLB upon a PTW update.


\subsection{Sstc Extension} \label{subsec:sstc}

Timer registers are exposed as MMIO registers (\textit{mtime}); however, the Sstc specification defines that \textit{stimecmp and vstimecmp} are hart CSRs. Thus, we exposed the time value to each hart via connection to the CLINT. We also added the ability to enable and disable this extension at S/HS-mode and VS-mode via \textit{menvcfg.STCE} and \textit{henvcfg.STCE} bits, respectively. For example, when \textit{menvcfg.stce} is 0, an S-mode access to stimecmp will trigger an illegal instruction. The same happens for VS-mode, when \textit{henvcfg.stce} is 0 (throwing a virtual illegal instruction exception). The Sstc extension does not break legacy timer implementations, as software that does not support the Sstc extension can still use the standard SBI interface to set up a timer. 

\section{Design Space Exploration: Evaluation} \label{sec:design-space-exploration}

In this section, we discuss the conducted design space exploration (DSE) evaluation. The system under evaluation, the DSE configuration, the followed methodology, as well as benchmarks and metrics are described below. Each subsection then focuses in assessing the functional performance speedup per the configuration of each specific module: (i) L1 TLB (Section \ref{subsec:l1-tlb-cap}); (ii) GTLB (Section \ref{subsec:gtlb_perf}); (iii) L2 TLB (Section \ref{subsec:l2tbl}); and (iv) Sstc (Section \ref{subsec:sstc_perf}).

\mypara{System and Tools.} We ran the experiments on a CVA6 single-core SoC featuring a 32KiB DCache and 16KiB ICache, targeting the Genesys2 FPGA at 100MHz. The software stack encompasses the (i) OpenSBI (version 1.0) (ii) Bao hypervisor (version 0.1), and Linux (version 5.17). We compiled Bao using SiFive GCC version 10.1.0 for riscv64 baremetal targets and OpenSBI and Linux using GCC version 11.1.0 for riscv64 Linux targets. We used Vivado version 2020.2 to synthesize the design and assess the impact on FPGA hardware resources.

\mypara{DSE Configurations.} The DSE configurations are summarized in Table \ref{table:dse-configs}. For each module, we selected the parameters we wanted to explore and their respective configurations. For instance, for the L1 TLB we fixed the number of entries as the design parameter and 16, 32, and 64 as possible configurations. Taking all modules, parameters, and configurations into consideration, it is possible to achieve up to 288 different combinations. Due to the existing limitations in terms of time and space, we carefully selected and evaluated 23 out of these 288 configurations.

\begin{table}[]
\caption{Design space exploration configurations.}
\centering
\begin{tabular}{l|l|ccc}
\hline
{\color[HTML]{000000} }                                              & {\color[HTML]{000000} }                                          & \multicolumn{3}{c}{{\color[HTML]{000000} \textbf{Configuration}}}                                                                             \\ \cline{3-5} 
\multirow{-2}{*}{{\color[HTML]{000000} \textbf{Module}}}             & \multirow{-2}{*}{{\color[HTML]{000000} \textbf{Parameter}}}      & \multicolumn{1}{c|}{{\color[HTML]{000000} \#1}}       & \multicolumn{1}{c|}{{\color[HTML]{000000} \#2}}       & {\color[HTML]{000000} \#3}          \\ \hline
{\color[HTML]{000000} L1 TLB}                                        & size entries                                                     & \multicolumn{1}{c|}{{\color[HTML]{000000} 16}}      & \multicolumn{1}{c|}{{\color[HTML]{000000} 32}}      & {\color[HTML]{000000} 64}         \\ \hline
{\color[HTML]{000000} GTLB}                                          & size entries                                                     & \multicolumn{1}{c|}{{\color[HTML]{000000} 8}}       & \multicolumn{1}{c|}{{\color[HTML]{000000} 16}}      & {\color[HTML]{000000} \----} \\ \hline
\multicolumn{1}{c|}{{\color[HTML]{000000} }}                         & page size                                                        & \multicolumn{1}{c|}{{\color[HTML]{000000} 4KiB}}    & \multicolumn{1}{c|}{{\color[HTML]{000000} 2MiB}}    & {\color[HTML]{000000} 4KiB+2MiB}  \\ \cline{2-5} 
\multicolumn{1}{c|}{{\color[HTML]{000000} }}                         & associativity                                                    & \multicolumn{1}{c|}{{\color[HTML]{000000} 4}}       & \multicolumn{1}{c|}{{\color[HTML]{000000} 8}}       & {\color[HTML]{000000} \----} \\ \cline{2-5} 
\multicolumn{1}{c|}{{\color[HTML]{000000} }}                         & \begin{tabular}[c]{@{}l@{}}size entries \\ for 4KiB\end{tabular} & \multicolumn{1}{c|}{{\color[HTML]{000000} 128}}     & \multicolumn{1}{c|}{{\color[HTML]{000000} 256}}     & {\color[HTML]{000000} \----} \\ \cline{2-5} 
\multicolumn{1}{c|}{\multirow{-4}{*}{{\color[HTML]{000000} L2 TLB}}} & \begin{tabular}[c]{@{}l@{}}size entries\\ for 2MiB\end{tabular}  & \multicolumn{1}{c|}{{\color[HTML]{000000} 32}}      & \multicolumn{1}{c|}{{\color[HTML]{000000} 64}}      & {\color[HTML]{000000} \----} \\ \hline
{\color[HTML]{000000} SSTC}                                          & status                                                           & \multicolumn{1}{c|}{{\color[HTML]{000000} enabled}} & \multicolumn{1}{c|}{{\color[HTML]{000000} disable}} & {\color[HTML]{000000} \----} \\ \hline
\end{tabular}
\label{table:dse-configs}
\vspace{-5mm}
\end{table}

\mypara{Methodology.} We focus the DSE evaluation on functional performance (execution cycles). We collected 1000 samples and computed the average of 95 percentile values to remove any outliers caused by some sporadic Linux interference. We normalized the results to the reference baseline execution (higher values translate to higher performance speedup) and added the absolute execution time on top of the baseline bar. Our baseline hardware implementation corresponds to a system with the CVA6 with hardware virtualization support but without the proposed microarchitectural extensions (e.g., GTLB, L2 TLB). We have also collected the post-synthesis hardware utilization; however, we omit the results due to lack of space (although we may occasionally refer to them during the discussion of the results in this section).

\mypara{Benchmarks.} We used the Mibench Embedded Benchmark Suite. \bs{The Mibench is an embedded application benchmark widely used in mixed-critically systems to assess the performance \cite{sa2021,martins2023}.} The Mibench incorporates a set of 35 application benchmarks grouped into six categories, targeting different embedded market segments. We focus our evaluation on the automotive subset. The automotive suite encompasses 3 high memory-intensive benchmarks (\textit{qsort}, \textit{susan corners} and \textit{susan edges}) that exercise many components across the memory hierarchy (e.g. MMU, cache, and memory controller). \bs{To complement our evaluation, we ran San Diego Vision Benchmark \cite{venkata2009} in the selected designs for the PPA analysis. San Diego Vision is a computer vision benchmark widely used to evaluate embedded and mixed-criticality systems \cite{kloda2023,gracioli2019}. It runs a full range of vision applications (e.g., motion tracking (\textit{tracking})) using multiple datasets with different sizes.  Instructions to run our experiments with Mibench and San Diego Vision can be found here\footnote{https://github.com/ninolomata/bao-cva6-guide/tree/cva6-evaluation}.}
\vspace{-5mm}
\begin{figure*}[t]
    \centering
    \includegraphics[width=0.9\textwidth,clip,trim=0cm 3cm 0cm 0cm]{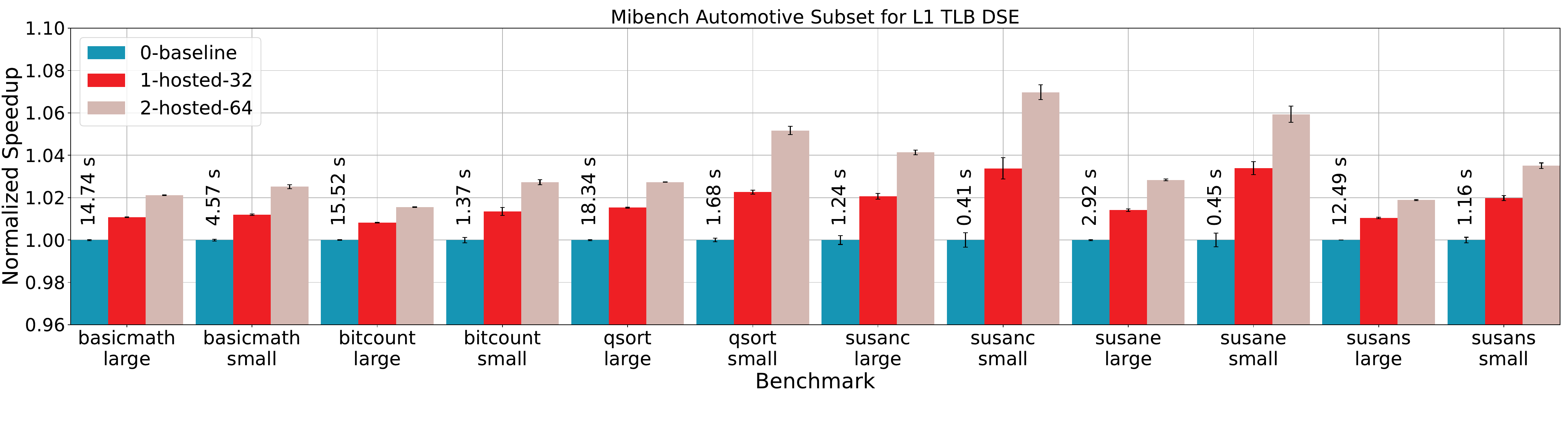}
    \caption{\bs{Mibench results for L1 TLB design space exploration evaluation.}}
    \label{fig:mibench_l1_tlb}
    \vspace{-2mm}
\end{figure*}

\subsection{L1 TLB} \label{subsec:l1-tlb-cap} 

In this subsection, we evaluate the functional performance for a different number of L1 TLB entries, i.e., 16, 32, and 64.

\begin{figure*}[!t]
    \centering
    \includegraphics[width=0.9\textwidth,clip,trim=0cm 3cm 0cm 0cm]{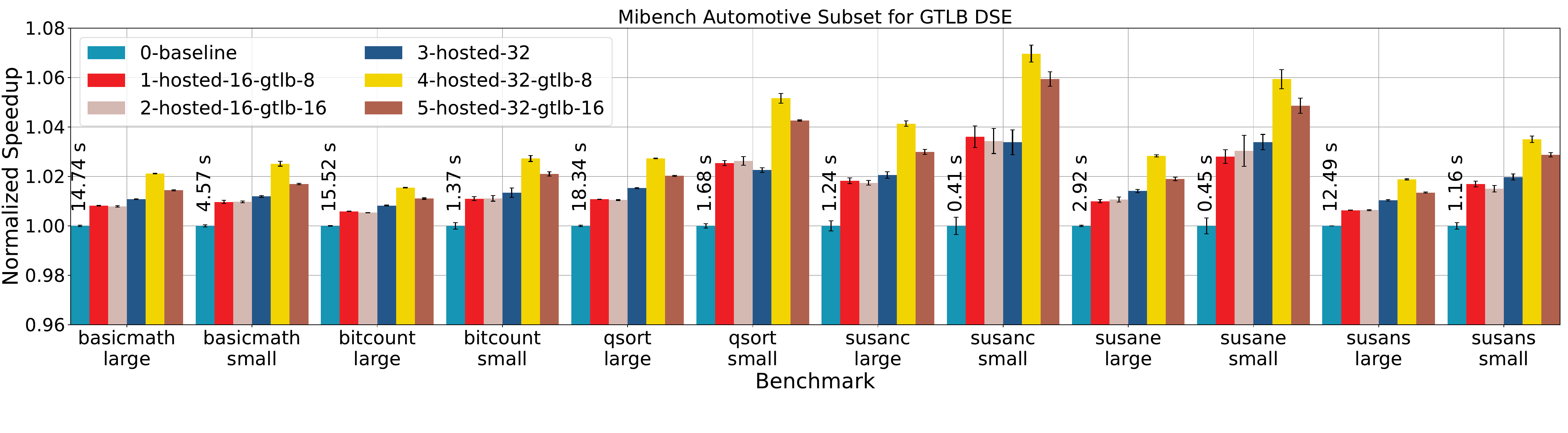}
    \caption{\bs{Mibench results for GTLB design space exploration evaluation.}}
    \label{fig:mibench_gtlb}
    \vspace{-5mm}
\end{figure*}

\mypara{L1 TLB Setup.} To assess the L1 TLB functional performance speedup, we ran the full set of benchmarks for three different setups: (i) Linux virtual/hosted execution for the baseline \textit{cva6-16} (\textit{baseline}); (ii) Linux virtual/hosted execution for the \textit{cva6-32} (\textit{hosted-32}); and (iii) Linux virtual/hosted execution for the \textit{cva6-64} (\textit{hosted-64}). 

\mypara{L1 TLB Performance Speedup.} Figure \ref{fig:mibench_l1_tlb} shows the assessed results. All results are normalized to the \textit{baseline} execution. Several conclusions can be drawn. Firstly, as expected, the \textit{hosted-64} is the best-case scenario with an average performance speedup increase of 3.6\%. We can observe a maximum speedup of 7\% in the \textit{susanc} (small) benchmark and a minimum speedup of 2\% in the \textit{bitcount} (large) benchmark. However, although not explicitly shown in this paper, to achieve these results there is an associated impact of 50\% increase in the FPGA resources. \bs{We expected these results because the L1 TLB is a fully-associative TLB implemented as a full combinational circuit, and the CVA6 design is not optimized for this particular FPGA architecture.} Secondly, the \textit{hosted-32} configuration increases the performance by a minimum of 1\% (e.g., basicmath large) and a maximum of 4\% (e.g., \textit{susanc} small), at a cost of about 15\%-17\% in the area (FPGA). Finally, we can conclude that increasing the CVA6 L1 TLB size to 32 entries presents the most reasonable trade-off between functional performance speedup and hardware cost.

\subsection{GTLB} \label{subsec:gtlb_perf}

In this subsection, we assess the GTLB impact on functional performance. We evaluate this for a different number of GTLB entries (8 and 16) and L1 TLB entries (16 and 32).

\mypara{GTLB Setup.} To assess the GTLB functional performance speedup, we ran the full set of benchmarks for seven distinct setups: (i) Linux virtual/hosted execution for the baseline \textit{cva6-16} (\textit{baseline}); (ii) Linux virtual/hosted execution for the \textit{cva6-32} (\textit{hosted-32}); (iii) Linux virtual/hosted execution for the \textit{cva6-16-gtlb-8} (\textit{hosted-16-gtlb-8}); (iv) Linux virtual/hosted execution for the \textit{cva6-32-gtlb-8} (\textit{hosted-32-gtlb-8}); (v) Linux virtual/hosted execution for the \textit{cva6-16-gtlb-16} (\textit{hosted-16-gtlb-16}); and (vi) Linux virtual/hosted execution for the \textit{cva6-32-gtlb-16} (\textit{hosted-32-gtlb-16}).

\mypara{GTLB Performance Speedup.} Mibench results are depicted in Figure \ref{fig:mibench_gtlb}. We normalized results to \textit{baseline} execution. We highlight a set of takeaways. Firstly, the \textit{hosted-16-gtlb-8} and \textit{hosted-16-gtlb-16} scenarios have similar results across all benchmarks with an average percentage performance speedup of about 2\%; therefore, there is no significant improvement from configuring the GTLB with 16 entries over 8 when the L1 TLB has 16 entries. Secondly, the \textit{hosted-32-gtlb-8} setup presents the best performance, i.e., a maximum performance increase of about 7\% for the \textit{susanc} (small) benchmark; however, at a non-negligible overall cost of 20\% in the hardware resources. Surprisingly, the hosted execution with 32 entries and a GTLB with 16 entries (\textit{hosted-32-gtlb-16}) perform slightly worst compared with an 8 entries GTLB (\textit{hosted-32-gtlb-8}). \bs{We have also collected microarchitectural hardware events on the CVA6. We noticed a slight increase in data cache misses for the \textit{hosted-32-gtlb-16} compared to the \textit{hosted-32-gtlb-8} of roughly 10\% in some benchmarks (e.g., \textit{susanc-large}), which explain why the GTLB with 16 entries performs worse than the GTLB with 8 entries.} For instance, for the \textit{susane} (large) benchmark the \textit{hosted-16-gtlb-8} setup achieves a 6\% performance increase while the \textit{hosted-16-gtlb-16} only 5\%. Finally, the \textit{hosted-32} achieves a performance increase in line with \textit{hosted-16-gtlb8} and \textit{hosted-16-gtlb16} configurations.

\subsection{L2 TLB} \label{subsec:l2tbl}

\begin{table}[!t]
\caption{L2 TLB design space exploration configurations.}
\begin{tabular}{l|c|l|l}
\hline
{\color[HTML]{000000} \textbf{Configuration}} & {\color[HTML]{000000} \textbf{L1 TLB}} & \multicolumn{1}{c|}{{\color[HTML]{000000} \textbf{GTLB}}} & \multicolumn{1}{c}{{\color[HTML]{000000} \textbf{L2 TLB}}}                                                                \\ \hline
{\color[HTML]{000000} cva6-16}           & {\color[HTML]{000000} 16 entries}      & \multicolumn{1}{c|}{\----}     & \multicolumn{1}{c}{\----}                                                              \\ \hline
{\color[HTML]{000000} cva6-16-gtlb8}           & {\color[HTML]{000000} 16 entries}      & \multicolumn{1}{c|}{{\color[HTML]{000000} 8 entries}}     & \multicolumn{1}{c}{\----}                                                              \\ \hline
{\color[HTML]{000000} cva6-16-l2-1}           & {\color[HTML]{000000} 16 entries}      & \multicolumn{1}{c|}{{\color[HTML]{000000} 8 entries}}     & \multicolumn{1}{c}{{\color[HTML]{000000} 4KiB - 128, 4-ways}}                                                             \\ \hline
{\color[HTML]{000000} cva6-16-l2-2}           & {\color[HTML]{000000} 16 entries}      & \multicolumn{1}{c|}{{\color[HTML]{000000} 8 entries}}     & \multicolumn{1}{c}{{\color[HTML]{000000} 2MiB - 32, 4-ways}}                                                              \\ \hline
{\color[HTML]{000000} cva6-16-l2-3}           & {\color[HTML]{000000} 16 entries}      & \multicolumn{1}{c|}{{\color[HTML]{000000} 8 entries}}     & \multicolumn{1}{c}{{\color[HTML]{000000} \begin{tabular}[c]{@{}c@{}}4KiB - 128, 4-ways\\ 2MiB - 32, 4-ways\end{tabular}}} \\ \hline
{\color[HTML]{000000} cva6-32-gtlb8}           & {\color[HTML]{000000} 16 entries}      & \multicolumn{1}{c|}{{\color[HTML]{000000} 8 entries}}     & \multicolumn{1}{c}{\----}                                                              \\ \hline
{\color[HTML]{000000} cva6-32-l2-1}           & {\color[HTML]{000000} 32 entries}      & \multicolumn{1}{c|}{{\color[HTML]{000000} 8 entries}}     & {\color[HTML]{000000} 4KiB - 128, 4-ways}                                                                                 \\ \hline
cva6-32-l2-2                                  & 32 entries                             & 8 entries                                                 & 4KiB - 128, 8-ways                                                                                                        \\ \hline
cva6-32-l2-3                                  & 32 entries                             & 8 entries                                                 & 4KiB - 256, 4-ways                                                                                                        \\ \hline
cva6-32-l2-4                                  & 32 entries                             & 8 entries                                                 & 4KiB - 256, 8-ways                                                                                                        \\ \hline
cva6-32-l2-5                                  & 32 entries                             & 8 entries                                                 & 2MiB - 32, 4-ways                                                                                                         \\ \hline
cva6-32-l2-6                                  & 32 entries                             & 8 entries                                                 & 2MiB - 32, 8-ways                                                                                                         \\ \hline
cva6-32-l2-7                                  & 32 entries                             & 8 entries                                                 & 2MiB - 64, 4-ways                                                                                                         \\ \hline
cva6-32-l2-8                                  & 32 entries                             & 8 entries                                                 & 2MiB - 64, 8-ways                                                                                                         \\ \hline
cva6-32-l2-9                                  & 32 entries                             & 8 entries                                                 & \multicolumn{1}{c}{\begin{tabular}[c]{@{}c@{}}4KiB - 128, 4-ways\\ 2MiB - 32, 4-ways\end{tabular}}                        \\ \hline
cva6-32-l2-10                                 & 32 entries                             & 8 entries                                                 & \multicolumn{1}{c}{\begin{tabular}[c]{@{}c@{}}4KiB - 256, 4-ways\\ 2MiB - 64, 4-ways\end{tabular}}                        \\ \hline
\end{tabular}
\label{table:l2-tlb-configs}
\vspace{-5mm}
\end{table}

\begin{figure*}[!t]
    \centering
    \includegraphics[width=1\textwidth,clip,trim=0cm 3cm 0cm 0cm]{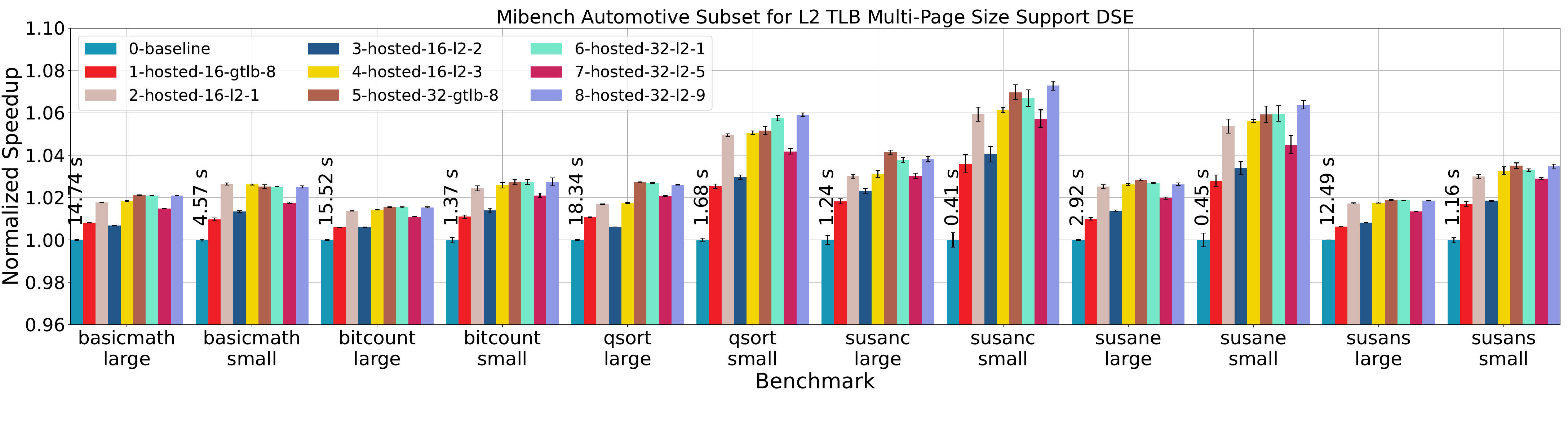}
    \caption{\bs{Mibench results for L2 TLB multi-page size support design space exploration evaluation.}}
    \label{fig:mibench_l2_tlb_size}
    \vspace{-2mm}
\end{figure*}

\begin{figure*}[!t]
    \centering
    \includegraphics[width=1\textwidth,clip,trim=0cm 3cm 0cm 0cm]{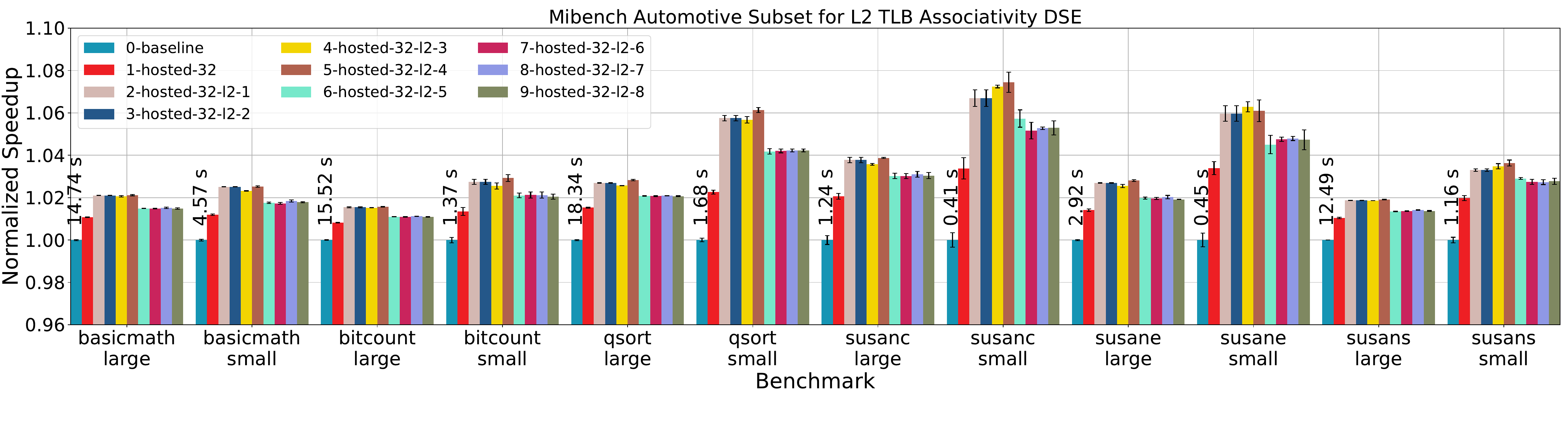}
    \caption{\bs{Mibench results for L2 TLB associativity design space exploration evaluation.}}
    \label{fig:mibench_l2_tlb_ways}
    \vspace{-5mm}
\end{figure*}

In this subsection, we assess the L2 TLB impact on functional performance. We conduct several experiments focusing on two main L2 TLB configuration parameters: (i) multi-page size support (4KiB or 2MiB or both) and (ii) TLB associativity. Furthermore, we also evaluated the L2 TLB impact in combination with different L1 TLB entries (16 and 32) and the GTLB with 8 entries. Table \ref{table:l2-tlb-configs} summarizes the design configurations.

\mypara{Multi-page Size Support Setup.} To assess the performance speedup for the L2 TLB multi-page size, we have carried out a set of experiments to measure the impact of: (i) 4KiB page size (configurations \textit{cva6-16-l2-1},\textit{cva6-32-l2-1}); (ii) 2MiB page size (configurations \textit{cva6-16-l2-2}, \textit{cva6-32-l2-5}); (iii) both 4KiB and 2MiB page sizes (configurations \textit{cva6-16-l2-3}, \textit{cva6-32-l2-9}); and (iv) increase the L1 TLB capacity. For each configuration, we run the Mibench benchmarks for nine different scenarios: (i) Linux virtual/hosted execution for the baseline \textit{cva6-l1-16} (\textit{baseline}); (ii) Linux native execution for the \textit{cva6-16} (\textit{bare}); (ii) Linux virtual/hosted execution for the \textit{cva6-16-gtlb8} (\textit{hosted-16-gtlb8}); (iii) Linux virtual/hosted execution for the \textit{cva6-16-l2-1} (\textit{hosted-16-l2-1}); (iv) Linux virtual/hosted execution for the \textit{cva6-16-l2-2} (\textit{hosted-16-l2-2}); (v) Linux virtual/hosted execution for the \textit{cva6-16-l2-3} (\textit{hosted-16-l2-3}); (vi) Linux virtual/hosted execution for the \textit{cva6-32-gtlb8} (\textit{hosted-32-gtlb8}); (vii) Linux virtual/hosted execution for the \textit{cva6-32-l2-1} (\textit{hosted-32-l2-1}); (viii) Linux virtual/hosted execution for the \textit{cva6-32-l2-5} (\textit{hosted-32-l2-5}); and (ix) Linux virtual/hosted execution for the \textit{cva6-32-l2-9} (\textit{hosted-32-l2-9}). 

\mypara{Multi-page Size Support Performance.} Mibench results are depicted in Figure \ref{fig:mibench_l2_tlb_size}. All results were normalized to \textit{baseline} execution. Based on Figure \ref{fig:mibench_l2_tlb_size}, we can extract several conclusions. First, configurations that support only 2MiB page size, i.e., \textit{hosted-16-l2-2} and \textit{hosted-16-l2-5}, present little or almost no performance improvement compared with the hardware configurations including only the GTLB (i.e., \textit{hosted-16-gtlb-8} and \textit{hosted-16-gtlb-8}). Second, supporting 4KiB page sizes causes a noticeable performance speedup for the L1 TLB with 16 entries, especially in memory-intensive benchmarks, such as \textit{qsort} (small), \textit{susanc} (small) and \textit{susane} (small). The main reason that justifies this improvement lies in the fact that Bao is configured to use 2MiB superpages; however, the Linux VM uses mostly 4KiB page size, i.e., most of the translations are 4KiB. From a different perspective, we observe similar results for the \textit{hosted-16-l2-1} and \textit{hosted-32-gtlb-8}, with few exceptions in some benchmarks (e.g., \textit{susane} (large), \textit{susanc} (large) and \textit{qsort} (large)). Moreover, for an L1 TLB configured with 32 entries, the average performance increase is roughly equal in less memory-intensive benchmarks (e.g. \textit{basicmath} and \textit{bitcount}). This is explained by the reduced number of L1 misses, due to its larger capacity which leads to fewer requests to the L2 TLB. Finally, we observe minimal speedup improvements on several benchmarks (e.g., \textit{susans}) when supporting both 4KiB and 2MiB (\textit{hosted-16-l2-3} and \textit{hosted-32-l2-9}).

\begin{figure*}[!t]
    \centering
    \includegraphics[width=1\textwidth,clip,trim=0cm 3cm 0cm 0cm]{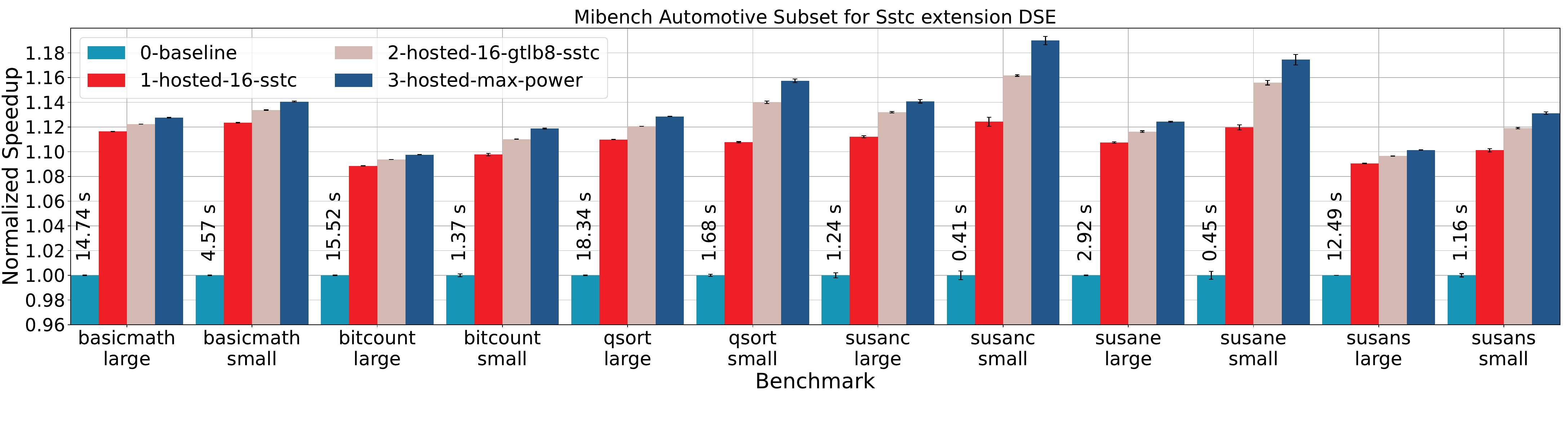}
    \caption{\bs{Mibench results for Sstc extension design space exploration evaluation.}}
    \label{fig:mibench_sstc}
    \vspace{-2mm}
\end{figure*}

\mypara{TLB associativity Setup.} To evaluate the performance speedup for the L2 TLB associativity, we select eight designs from previous experiments and modified the following parameters: (i) the number of 4KiB entries (128 or 256) and 2MiB (32 or 64) entries for the L2 TLB; and (ii) the L2 TLB associativity, by configuring the L2 TLB in the 4-way or 8-way scheme. For each configuration, we have run the selected benchmarks for ten configurations: (i) Linux virtual/hosted execution for the baseline \textit{cva6-l1-16} (\textit{baseline}); (ii) Linux virtual/hosted execution for the \textit{cva6-32} (\textit{hosted-32}); (iii) Linux virtual/hosted execution for the \textit{cva6-32-l2-1} (\textit{hosted-32-l2-1}); (iv) Linux virtual/hosted execution for the \textit{cva6-32-l2-2} (\textit{hosted-32-l2-2}); (v) Linux virtual/hosted execution for the \textit{cva6-32-l2-3} (\textit{hosted-32-l2-3}); (vi) Linux virtual/hosted execution for the \textit{cva6-32-l2-4} (\textit{hosted-32-l2-4}); (vii) Linux virtual/hosted execution for the \textit{cva6-32-l2-5} (\textit{hosted-32-l2-5}); (viii) Linux virtual/hosted execution for the \textit{cva6-32-l2-6} (\textit{hosted-32-l2-6}); (ix) Linux virtual/hosted execution for the \textit{cva6-32-l2-7} (\textit{hosted-32-l2-7}); and (x) Linux virtual/hosted execution for the \textit{cva6-32-l2-7} (\textit{hosted-32-l2-7}).

\mypara{TLB associativity Performance.} The results in Figure \ref{fig:mibench_l2_tlb_ways} demonstrate that there is no significant improvement in modifying the number of sets and ways in 2MiB page size configurations (i.e., \textit{hosted-32-l2-5} to \textit{hosted-32-l2-8}). Additionally, we found that: (i) increasing the associativity from 4 to 8 or doubling the capacity in low memory-intensive benchmarks (e.g., \textit{bitcount}) had little impact on functional performance; and (ii) memory-intensive benchmarks (e.g., \textit{susanc} (small)) normaly present a speedup when increasing the number of entries for the same associativity (although with a larger standard deviation). For instance, in the \textit{susanc} (small) benchmark, we observer a performance speedup from 6\% to 7\% when the page support was 4KiB with 128 and 256 capacity organized into a 4-way per set (\textit{hosted-32-l2-1} and \textit{hosted-32-l2-3}).

\subsection{Sstc Extension} \label{subsec:sstc_perf}

\begin{table}[]
\caption{SSTC design space exploration configurations.}
\begin{tabular}{l|c|c|c|l}
\hline
\textbf{Configuration} & \textbf{L1 TLB} & \textbf{GTLB} & \textbf{L2 TLB} & \textbf{SSTC}                                                                                \\ \hline
cva6-sstc            & 16 & \---- & \---- & enabled \\ \hline
cva6-sstc-gtlb8            & 16 & 8 & \---- & enabled \\ \hline
cva6-max-power           & 16 & 8  & \begin{tabular}[c]{@{}c@{}}4KiB - 128, 4-ways\\ 2MiB - 32, 4-ways\end{tabular} & enabled \\ \hline
\end{tabular}
\label{table:sstc-configs}
\vspace{-5mm}
\end{table}

In this subsection, we assess the Sstc extension impact on performance. We assess the Sstc performance speedup for a few configurations exercised in the former subsections. Table \ref{table:sstc-configs} summarizes the design configurations.

\begin{figure*}[!t]
    \centering
    \includegraphics[width=1\textwidth,clip,trim=0cm 3cm 0cm 0cm]{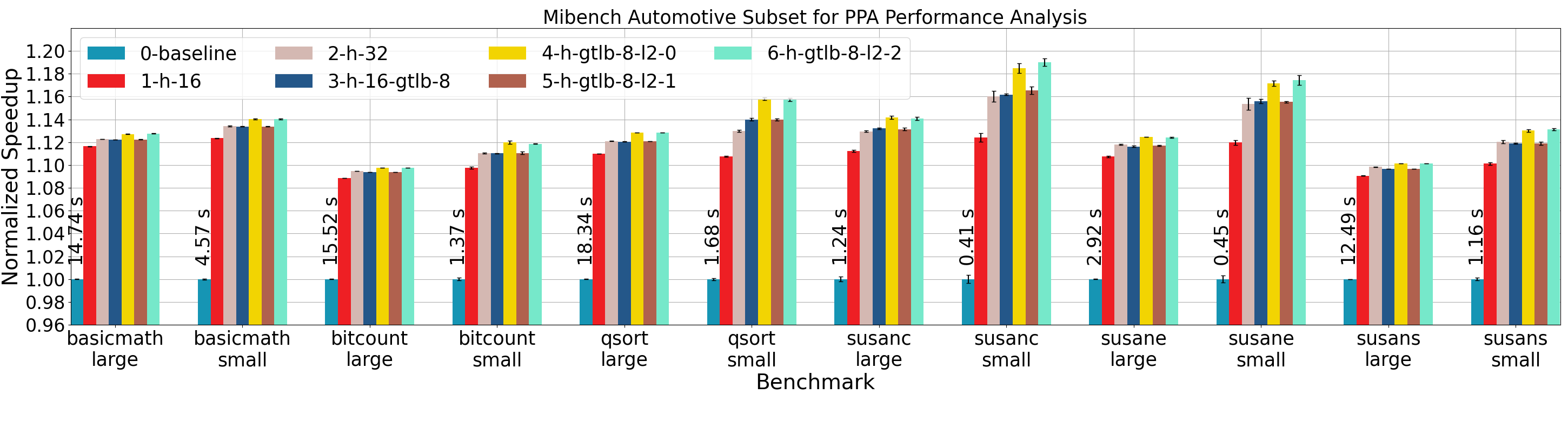}
    \caption{\bs{Summary of MiBench functional performance results for selected configurations.}}
    \label{fig:performance_PPA}
    \vspace{-5mm}
\end{figure*}

\begin{figure*}[!t]
    \centering
    \includegraphics[width=1\textwidth,clip,trim=0cm 3cm 0cm 0cm]{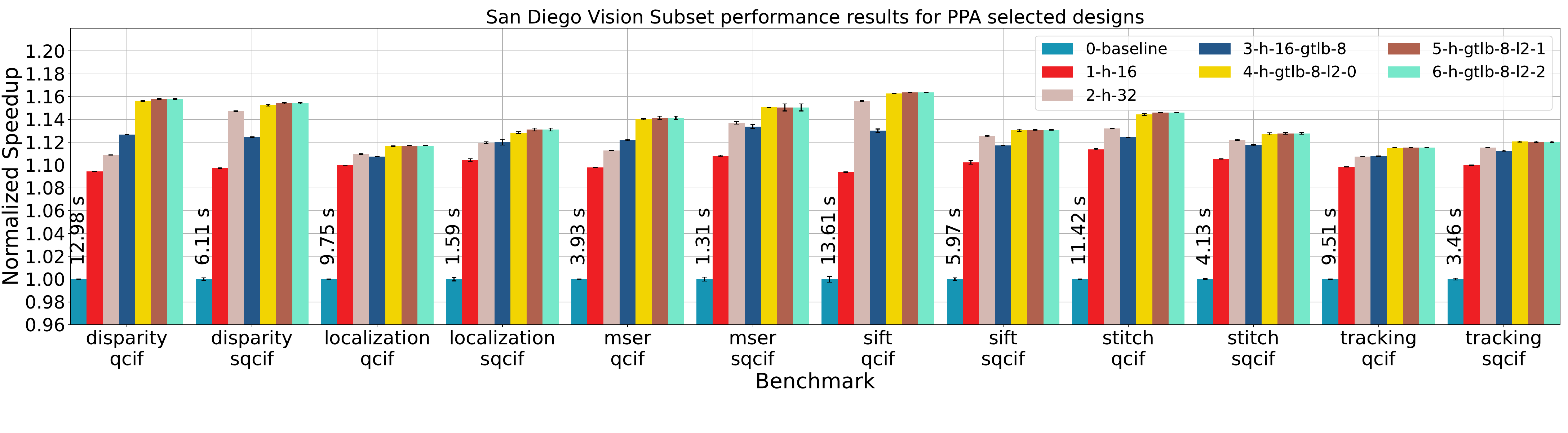}
    \caption{\bs{San Diego Vision Benchmark functional performance results for PPA analysis selected configurations.}}
    \label{fig:performance_PPA_SDV}
    \vspace{-5mm}
\end{figure*}

\mypara{Sstc Extension Setup.} To assess the Sstc extension performance speedup, we ran the full set of benchmarks for four different setups: (i) Linux virtual/hosted execution for the baseline \textit{cva6-16} (\textit{baseline}); (ii) Linux virtual/hosted execution for \textit{cva6-sstc} (\textit{hosted-16-sstc}); (iii) Linux virtual/hosted execution for \textit{cva6-gtlb8-sstc} (\textit{hosted-16-gtlb8-sstc}); and (iv) Linux virtual/hosted execution for \textit{cva6-max-power} (\textit{hosted-max-power-sstc}).

\mypara{Sstc Extension Performance.} The results depicted in Figure \ref{fig:mibench_sstc} show that for the \textit{hosted-16-sstc} scenario, there is a significant performance speedup. For example, in the memory-intensive \textit{qsort} benchmark, the performance speedup is 10.5\%. Timer virtualization is a major cause of pressure on the MMU subsystem due to the frequent transitions between the Hypervisor and Guest. The "hosted-max-power" scenario performs the best, with an average performance increase of 12.6\%, ranging from a minimum of 10\% in the \textit{bitcount} benchmark to a maximum of 19\% in memory-intensive benchmarks such as the \textit{susanc}. Finally, in less memory-intensive benchmarks, e.g., \textit{basicmath} (large) and \textit{bitcount} (large), the \textit{hosted-16-gtlb8-sstc} scenario performs similarly to the \textit{hosted-max-power} scenario, but with significantly less impact on hardware resources. For example, in the \textit{basicmath} (large) benchmark, there is a negligible difference of 0.5 \% comparing the \textit{hosted-16-gtlb8-sstc} with the \textit{hosted-max-power} scenario.

\subsection{Designs for PPA Analysis Selection} \label{subsec:ppa_analy}

\bs{We selected six configurations for the PPA analysis (see Table \ref{tab:ppa_conf}) based on the functional performance setup results summarized in Figure \ref{fig:performance_PPA}. We also ran San Diego Vision Benchmark to evaluate the selected designs with different workloads.}

\mypara{\bs{San Diego Vision Workloads Results.}} \bs{We ran the San Diego Vision Benchmark for the configurations elected for the PPA analysis. The results depicted in Figure \ref{fig:performance_PPA_SDV} are in line with the results collected for the Mibench benchmark (Figure \ref{fig:performance_PPA}). First, for the \textit{h-16} scenario, there is an average performance speedup of 10\%. Second, the \textit{h-32} outperforms \textit{h-16-gtlb-8} in some benchmarks with few exceptions. For instance, for the \textit{disparity qcif}, there is a slight difference of 2\%. Third, \textit{h-16-gtlb-8} has reasonable results with an average performance increase of about 12\% (maximum of 14\% and minimum of 10\%). Finally, for the \textit{h-gtlb-8-l2-0}, \textit{h-gtlb-8-l2-1}, and \textit{h-gtlb-8-l2-2} configurations, the performance is identical, with a pattern ranging a maximum of 16\% and a minimum of 11\%.}

\section{Physical Implementation} \label{ppa-analysis}

\subsection{Methodology}

Based on the functional performance results discussed in Section \ref{sec:design-space-exploration} and summarized in Figure \ref{fig:performance_PPA}, we select six configurations to compare with the baseline implementation of CVA6 with hardware virtualization support. Table \ref{tab:ppa_conf} lists the hardware configurations under study. We implemented these configurations in 22 nm FDX technology from Global Foundries down to ready-for-silicon layout to reliably estimate their operating frequency, power, and area.

\begin{figure}[t!]
    \centering
    \includegraphics[clip,width=0.5\textwidth]{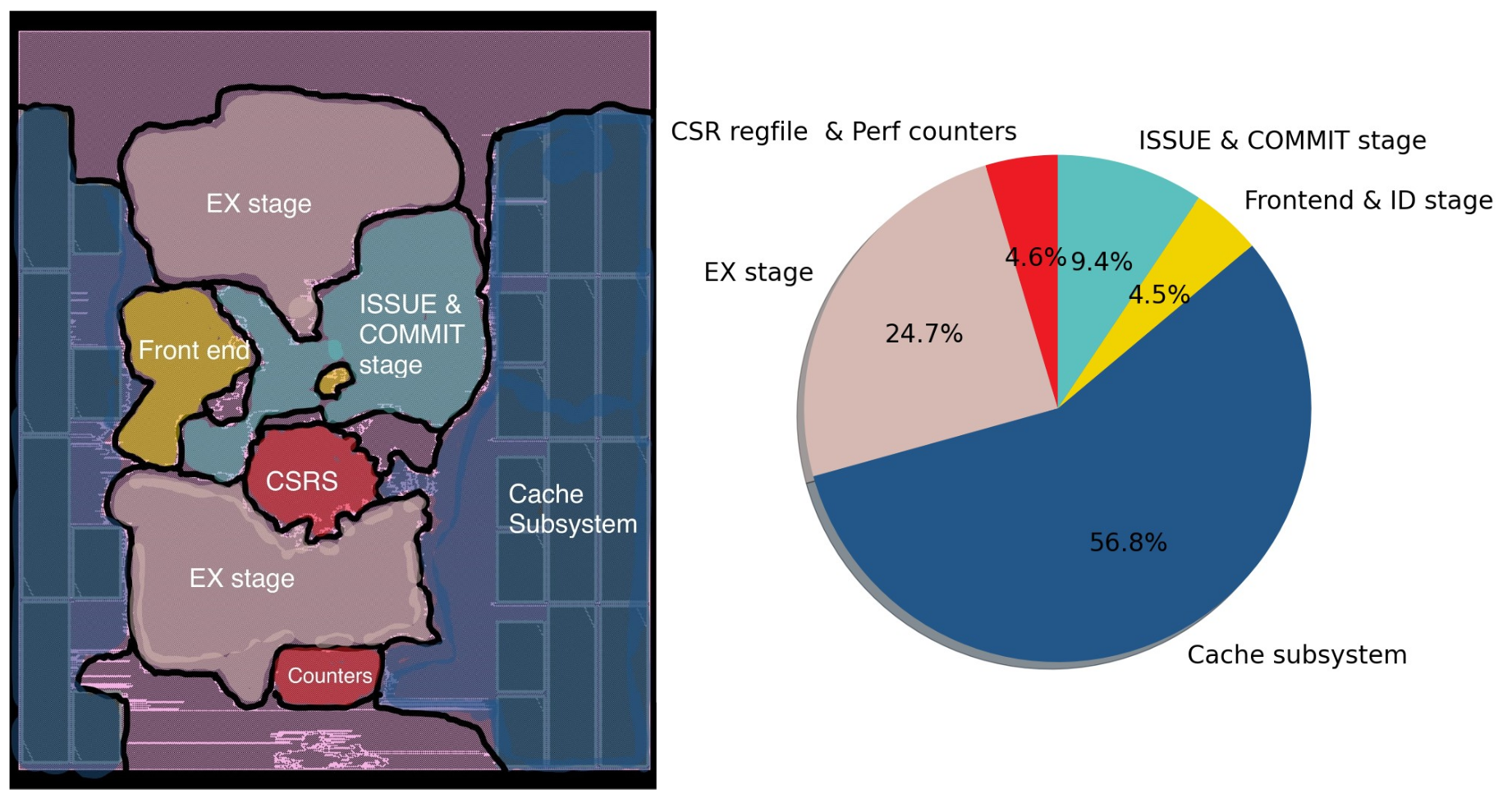}
    \caption{Baseline CVA6 layout 0.75mmx0.65mm}
    \label{fig:layout}
    \vspace{-5mm}
\end{figure}

To support the physical implementation and the PPA analysis, we used the following tools: (i) Synopsys Design Compiler 2019.03 to perform the physical synthesis; (ii) Cadence Innovus 2020.12 for the place \& route; (iii) Synopsys PrimeTime 2020.09 for the power analysis - extracting value change dump (VCD) traces on the post layout; and (iv) Siemens Questasim 10.7b to design the parasitics annotated netlist. Figure~\ref{fig:layout} shows the layout of CVA6, featuring an area smaller than 0.5mm$^2$. 

\begin{figure*}[!t]
    \centering
    \includegraphics[width=\textwidth,clip]{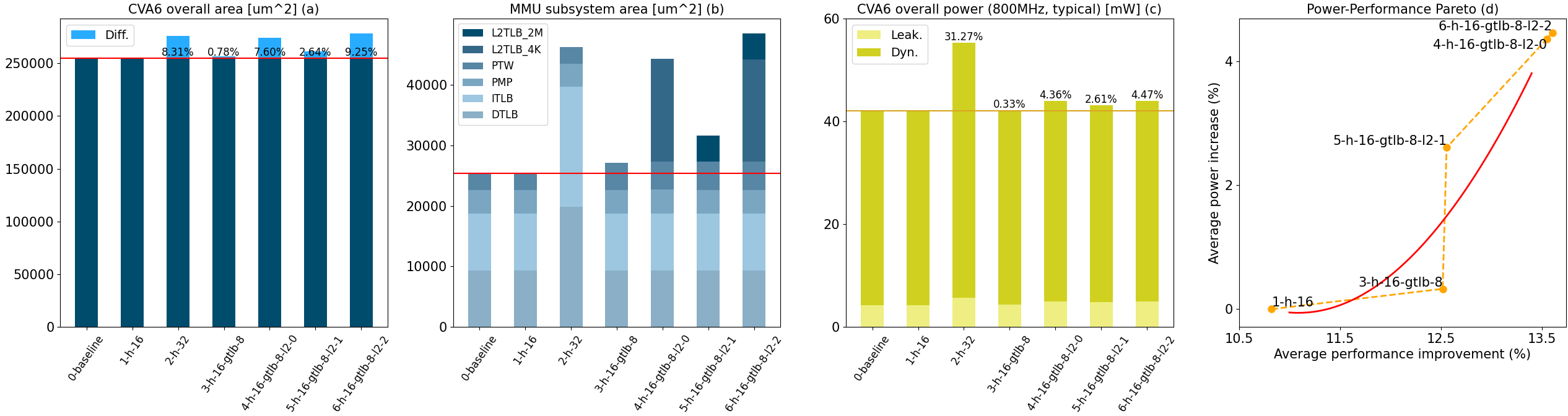}
    \caption{Area and Power results}
    \label{fig:PPA}
    \vspace{-5mm}
\end{figure*}

\begin{table}[!t]
    \centering
    \begin{tabular}{c|c|c|c|c|c} \hline
                         & SSTC       &  I/DTLB    &  8-entries & 4k L2      & 2MB L2 \\ 
                         & support    &  \#entries &  GTLB      & TLB        & TLB    \\ \hline
         0-vanilla       & x          &  16        & x          & x          & x      \\ \hline
         1-h-16          & \checkmark &  16        & x          & x          & x      \\ \hline
         2-h-32          & \checkmark &  32        & x          & x          & x      \\ \hline
         3-h-gtlb-8      & \checkmark &  16        & \checkmark & x          & x      \\ \hline
         4-h-gtlb-8-l2-0 & \checkmark &  16        & \checkmark & \checkmark & x      \\ \hline
         5-h-gtlb-8-l2-1 & \checkmark &  16        & \checkmark & x          & \checkmark \\ \hline
         6-h-gtlb-8-l2-2 & \checkmark &  16        & \checkmark & \checkmark & \checkmark \\ \hline
    \end{tabular}
    \caption{7 selected configurations for PPA analysis in GF22nm}
    \label{tab:ppa_conf} 
    \vspace{-10mm}
\end{table}

\vspace{2pt}

\subsection{Results}

We fix the target frequency to 800MHz in the worst corner (SSG corner at 0.72 V, -40/125 \degree C). All configurations manage to reach the target frequency. We then compare the area and power consumption while running a dense 16x16 FP matrix multiplication at 800MHz with warmed-up caches (TT corner, 25 \degree C). Leveraging the extracted power measurements and the functional performance on the Mibench benchmarks, we further obtain the relative energy efficiency.

\begin{figure*}[!t]
    \centering
    \includegraphics[width=1\textwidth,clip,trim=0cm 3cm 0cm 0cm]{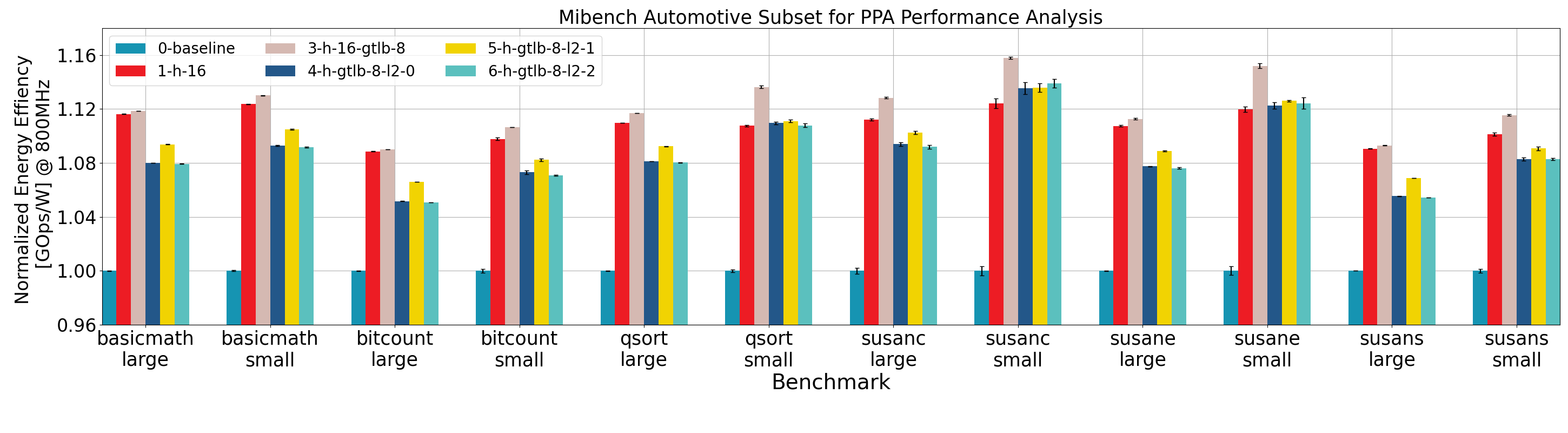}
    \caption{Energy efficiency results.}
    \label{fig:energy}
    \vspace{-5mm}
\end{figure*}

Figure \ref{fig:PPA} depicts the PPA results. Figure \ref{fig:PPA}(a) shows that the Sstc extension has negligible impact on the area. In fact, as expected, the MMU is the microarchitectural component with a higher impact on power and area. Figure \ref{fig:PPA}(b) highlights the MMU area comparison. The configuration with 32 entries doubles the ITLB and DTLB area, while the other configurations have little to no impact on the ITLB and DTLB; they, at most, add the GTLB and the L2-TLB modules on top of the existing MMU configuration. Figure \ref{fig:PPA}(c) shows the measured power consumption. We observe that increasing the number of L1 TLB entries from 16 to 32 (2-h-32) increases the power by 31\%, while the other configurations impact less than 5\% on power compared with the vanilla CVA6. Figure \ref{fig:energy} shows the relative energy efficiency on the MiBench benchmarks. The second configuration (2-h-32) is less energy efficient than the baseline since the performance gain ($\leq 15\%$) is smaller than the power increase ($\geq 30\%$). It is therefore excluded in the graph analysis. On the other hand, all the other configurations increase the energy efficiency up to 16\%.

Figure \ref{fig:PPA}(d) plots the measured power consumption of the energy-efficient configurations against the average performance improvement on the Mibench benchmarks. The SSTC extension alone brings an average performance improvement of around 10\% with a negligible power increase. At the same time, the explored MMU configurations can offer a clean trade-off between performance and power. The most expensive configuration can provide an extra performance of 4\% gain for a 4.47\% power increase. Lastly, the hardware configuration including the Sstc support and a GTLB with 8 entries (3-h-16-gtlb-8), is the most energy-efficient one, with the highest ratio between performance and power increase. 

In conclusion, we can argue that the hardware configuration with Sstc support and a GTLB with 8 entries (3-h-16-gtlb-8) is the optimal design point since we can achieve a functional performance speedup of up to 16\% (approx. 12.5\% on average) at the cost of 0.78\% in area and 0.33\% in power.

\section{Discussion}
\mypara{\bs{Performance Evaluation.}} \bs{We carried out an extensive functional performance evaluation for multiple FPGA design configurations in a single-core environment. As part of future work, we plan and encourage the evaluation of additional configurations and software stacks. Possible directions include (i) experimenting with other hypervisors like XVisor \cite{Patel2015} in different configurations (e.g., having two guests running in a single core); (ii) running other commercial benchmarks like SPEC17; and (iii) shifting to multi-core environments, e.g., available in frameworks such as OpenPiton \cite{balkind2016}.}

\mypara{System-level Virtualization.} The RISC-V ISA has reached a new state of maturity, and some of its previously existing gaps \cite{sa2021} (e.g., no standard interrupt controller with interrupt virtualization support) are already addressed. The Advanced Interrupt Architecture (AIA) \cite{hauser2022} specification has been ratified, and it provides preliminary support for interrupt virtualization using Message Signaled Interrupts (MSI). Hypervisors can leverage the AIA to reduce interrupt latency to guests and improve overall performance. As part of our ongoing efforts, we are working on providing an open-source reference implementation of the AIA IPs. The RISC-V I/O memory management unit (IOMMU) \footnote{https://github.com/riscv-non-isa/riscv-iommu} is also now ratified. Fundamentally, the IOMMU protects memory accesses from DMA-capable devices while guaranteeing isolation between VMs and the hypervisor. For instance, in Bao, all devices are assigned in a one-to-one scheme to a single guest using pass-through device mechanisms, which means Bao entirely relies on the IOMMU to do this. As part of our ongoing efforts, we are working on providing open-source reference implementations of the AIA IPs and the IOMMU IP.

\mypara{Memory Subsystems Improvements.} From an MMU microarchitectural point of view, there is ample room for improvement. We demonstrated that the CVA6 memory subsystem is the major cause of performance degradation, and further improvements are of utmost importance. For instance, we could redesign the CVA6 MMU to support TLB coalescing by augmenting the PTE to support NAPOT pages (Snapot extension). Although we have accelerated the PTW with dedicated TLB for the second stage, a few memory accesses to the memory hierarchy are still needed, i.e., side-effects are still expected at the cache level, and memory bandwidth at the system interconnect. One could explore using a PTE cache to store PTE entries close to the PTW and thus reduce the number of memory accesses and cache pollution. From the platform standpoint, RISC-V rich environment offers extensions that can benefit virtualized environments. We already discussed the standard cache management instructions, i.e., CMOs, defined in the Zicbom and Zicboz extensions. Hypervisors like Bao depend upon such instructions to implement partitioning mechanisms such as cache coloring, which otherwise would have to flush caches using firmware \textit{ecalls}. Finally, this work targeted a single-core platform with no Last-Level of Cache (LLC). Hence, the effectiveness of such mechanisms is yet to be proved in a multi-core environment where inter-core interference is visible throughout the memory hierarchy. An all-new class of security challenges is present, e.g., timing side-channels \cite{Ge2018}.

\section{Related Work}

The RISC-V hypervisor extension is a relatively new addition to the privileged architecture of the RISC-V ISA, ratified as of December 2021. Notwithstanding, from a hardware perspective, there are already a few open-source and commercial RISC-V cores with hardware virtualization support (or ongoing support): (i) the open-source Rocket core \cite{Asanovic2016}; (ii) the space-grade NOEL-V \cite{novel-v,jan202}, from Cobham Gaisler; (iii) the commercial SiFive P270, P500, and P600 series; (iv) the commercial Ventana Veryon V1; (v)  the commercial Dubhe from StarFive; (vi) the open-source Chromite from InCore Semiconductors; and (vii) the open-source Shakti core from IIT Madras.

\begin{table*}[]
\caption{CVA6 alike RISC-V Cores with Hypervisor Extension support features: \checkmark - supported;  X - not supported; and ? - no information available.}
    \center
\resizebox{\textwidth}{!}{
\begin{tabular}{l|c|c|c|c|c|c|c|c|c}
\hline
\textbf{Processor}                                        & \textbf{Pipeline}                                                       & \textbf{\begin{tabular}[c]{@{}c@{}}Priv.\\ Hyp.\end{tabular}} & \textbf{SSTC}       & \textbf{2D-MMU}                                                           & \textbf{L1 Cache}                                                                  & \textbf{L1 TLB}                                                                                                                                                           & \textbf{L2 TLB}                                                                                                                                          & \textbf{\begin{tabular}[c]{@{}c@{}}PTW \\ Opts\end{tabular}} & \textbf{Status} \\ \hline
Rocket                                                    & \begin{tabular}[c]{@{}c@{}}5-stage\\ in-order\end{tabular}              & v0.6                                                          & X                   & SV39x4                                                                    & \begin{tabular}[c]{@{}c@{}}32 KiB, \\ I/DCache\end{tabular}                                                    & \begin{tabular}[c]{@{}c@{}}set-assoc. 4KiB I/DTLB\\ or\\ full-assoc. superpages I/DTLB,\\ 4-entries (configurable)\end{tabular}                                    & \begin{tabular}[c]{@{}c@{}}dir.-mapped, (configurable size)\end{tabular}                                                                             & \begin{tabular}[c]{@{}c@{}}PTE \\ Cache\end{tabular}         & Done            \\ \hline
NOEL-V                                                    & \begin{tabular}[c]{@{}c@{}}7-stage\\ dual-issue\\ in-order\end{tabular} & v0.6                                                          & \checkmark          & \begin{tabular}[c]{@{}c@{}}SV32x2\\ SV39x4\end{tabular}                   & \begin{tabular}[c]{@{}c@{}}32 KiB \\ I/DCache\end{tabular}                         & \begin{tabular}[c]{@{}c@{}}I/DTLB,\\ (configurable)\end{tabular}                                                                                                          & X                                                                                                                                                        & hTLB                                                         & Done            \\ \hline
Chromite                                                  & \begin{tabular}[c]{@{}c@{}}6-stage\\ in-order\end{tabular}              & v1.0                                                          & ?                   & \begin{tabular}[c]{@{}c@{}}SV32x4\\ SV39x4\\ SV48x4\\ SV57x4\end{tabular} & \begin{tabular}[c]{@{}c@{}}32 KiB\\ I/DCache\end{tabular}                                                                                                          & \begin{tabular}[c]{@{}c@{}}set-assoc. split I/DTLB,\\ (configurable page sizes support)\\ or \\ full-assoc. I/DTLB, ?-entries (def. 4)\end{tabular} & X                                                                                                                                                        & X                                                            & WIP             \\ \hline
\begin{tabular}[c]{@{}l@{}}Shaktii\\ C Class\end{tabular} & \begin{tabular}[c]{@{}c@{}}5-stage\\ in-order\end{tabular}              & v1.0                                                          & ?                   & \begin{tabular}[c]{@{}c@{}}SV32x4\\ SV39x4\\ SV48x4\end{tabular}          & \begin{tabular}[c]{@{}c@{}}16 KiB\\ I/DCache\end{tabular}                                                                                                        & \begin{tabular}[c]{@{}c@{}}fully assoc. L1 I/DTLB,\\ 4-entries\end{tabular}                                                                                               & X                                                                                                                                                        & X                                                            & WIP             \\ \hline
\textbf{CVA6}                                             & \textbf{\begin{tabular}[c]{@{}c@{}}6-stage\\ in-order\end{tabular}}     & \textbf{v1.0}                                                 & \textbf{\checkmark} & \textbf{SV39x4}                                                           & \textbf{\begin{tabular}[c]{@{}c@{}}32 KiB\\ DCache,\\ 16KiB\\ ICache\end{tabular}}                                                                        & \textbf{\begin{tabular}[c]{@{}c@{}}full-assoc. I/DTLB, \\ 4-entries\end{tabular}}                                                                                         & \textbf{\begin{tabular}[c]{@{}c@{}}set assoc. 4KiB 128-256 entries, 4-8 ways\\ or/and\\ set assoc. 2MiB  32-64 entries, 4-8 ways\end{tabular}} & \textbf{GTLB}                                                & \textbf{Done}   \\ \hline
\end{tabular}
}
\label{tab:riscv-cores-hyp}
\vspace{-5mm}
\end{table*}

Table \ref{tab:riscv-cores-hyp} summarizes some information about the open-source cores, focusing on their microarchitectural and architectural features relevant to virtualization. Table \ref{tab:riscv-cores-hyp} shows that RISC-V cores with full support for the hypervisor extension (e.g., the Rocket core and NOEL-V) are only compliant with version 0.6 of the specification. The work described in this paper is already fully compliant with ratified version 1.0. Regarding the microarchitecture: (i) the Rocket core has some general MMU optimizations (e.g., PTE cache and L2 TLB), non of which are special targeting virtualization (like our GTLB); (ii) Chromite and Shaktii have no MMU optimizations for virtualization (and their support for virtualization is still in progress); and (iii) NOEL-V has a dedicated hypervisor TLB (hTLB). From a Cache hierarchy point of view, all cores have identical sizes of L1 Cache, except the CVA6 ICache, which is smaller (16KiB).

In terms of functional performance, our prior work on the Rocket Core \cite{sa2021} concluded an average of 2\% performance overhead due to the hardware virtualization support. Findings in this work are in line with the ones presented in \cite{sa2021} (however, with a higher penalty in performance due to the area-energy focus of the microarchitecture of the CVA6). Notwithstanding, we were able to optimize and extend the microarchitecture to significantly improve performance at a fraction of area and power. Results in terms of performance are not available for other cores, i.e., NOEL-C, Chromite, and Shaktii. From an energy and area breakdown perspective, none of these cores have made a complete public PPA evaluation. NOEL-V only reported a 5\% area increase for adding the hypervisor extension.

A large body of literature also covers techniques to optimize the memory subsystem. Some focus on optimizing the TLB \cite{chang_2007,pham2012,talluri1992,pham2014,cox2017,ryoo2017,Kandiraju2002,bhattacharjee2011,Bharadwaj2018}, while others aim at optimizing the PTW \cite{barr2010, bhargava2008}. For instance, in \cite{bhargava2008}, authors proposed a dedicated TLB on the PTW to skip the second translation stage and reduce the number of walk iterations. Our proposal for the GTLB structure is similar. Nevertheless, to the best of our knowledge, this is the first work to present a design space exploration evaluation and PPA analysis for MMU microarchitecture enhancements in the context of virtualization in RISC-V, fully supported by a publicly accessible open-source design \footnote{https://github.com/minho-pulp/cva6/tree/feature-hyp-ext-final}.

%
\section{Conclusion}

This article reports our work on hardware virtualization support in the RISC-V CVA6 core. We start by describing the baseline extension to the vanilla CVA6 to support virtualization and then focus on the design of a set of enhancements to the nested MMU. We first designed a dedicated G-Stage TLB to speedup the nested-PTW on TLB Miss. Then, we proposed an L2 TLB with multi-page size support (4KiB and 2MiB) and configurable size. We carried out a design space exploration evaluation on an FPGA platform to assess the impact on functional performance (execution cycles) and hardware. Then, based on the DSE evaluation, we elected a few hardware configurations and performed a PPA analysis based on 22nm FDX technology. Our analysis demonstrated several design points on the performance, power, and area trade-off. We were able to achieve a performance speedup of up to 19\% but at the cost of a 10\% area increase. We selected an optimal design configuration where we observed an average performance speedup of 12.5\% at a fraction of the area and power, i.e., 0.78\% and 0.33\%, respectively.

\section*{Acknowledgments}

This work has been supported by Technology Innovation Institute (TII), the European PILOT (EuroHPC JU, g.a. 101034126), the EPI SGA2 (EuroHPC JU, g.a. 101036168), and FCT – Fundação para a Ciência e Tecnologia within the R\&D Units Project Scope UIDB/00319/2020 and Scholarships Project Scope SFRH/BD/138660/2018 and SFRH/BD/07707/2021.

 \bibliographystyle{IEEEtran}
 \bibliography{main}

\end{document}